\documentclass[useAMS]{mn2e}
\usepackage{psfig,epsfig}
\usepackage{graphicx}

\begin{document}
\def\simlt{\mathrel{\rlap{\lower 3pt\hbox{$\sim$}}
        \raise 2.0pt\hbox{$<$}}}
\def\simgt{\mathrel{\rlap{\lower 3pt\hbox{$\sim$}}
        \raise 2.0pt\hbox{$>$}}}

\title[The PEP survey: clustering of infrared-selected galaxies and structure formation at $z\sim2$ in the GOODS South]{The PEP survey: 
clustering of infrared-selected galaxies and structure formation at $z\sim2$ in the GOODS South\thanks{Herschel is an ESA space observatory with science
instruments provided by European-led Principal Investigator consortia
and with important participation from NASA.}}
\author[Manuela Magliocchetti et al.]
{\parbox[t]\textwidth{M. Magliocchetti$^{1}$, P. Santini$^{2,3}$, G. Rodighiero$^{4}$, A. Grazian$^{2}$, H. Aussel$^{5}$, B. Altieri$^{6}$, 
P. Andreani$^{7}$, S. Berta$^{3}$, J. Cepa$^{8}$, H. Casta\~neda$^{8}$, A. Cimatti$^{9}$, E. Daddi$^{5}$, D. Elbaz$^{5}$, R. Genzel$^{7}$, 
C. Gruppioni$^{10}$, D. Lutz$^{3}$,  B. Magnelli$^{3}$, R. Maiolino$^{2}$, P. Popesso$^{3}$, A. Poglitsch$^{3}$, F. Pozzi$^{9}$,
 M. Sanchez-Portal$^{6}$, N.M.F\"orster Schreiber$^{3}$, E. Sturm$^{3}$, L. Tacconi$^{3}$, I. Valtchanov$^{6}$}\\
 \\
{\tt $^1$ INAF-IFSI, Via Fosso del Cavaliere 100, 00133, Roma, Italy}\\
{\tt $^2$ INAF-Osservatorio Astronomico di Roma, Via di Frascati 33, 00040 Monte Porzio Catone, Italy}\\
{\tt $^3$ Max Planck Institute f\"ur Extraterrestrische Physik (MPE), Postfach 1312,  D85741, Garching, Germany}\\
{\tt $^{4}$ Dipartimento di Astronomia, Universita' di Padova,  Vicolo dell' Osservatorio 3, 35122, Padova, Italy}\\
{\tt $^5$ CEA-Saclay, Service d'Astrophysique, F-91191, Gif-sur-Yvette, France}\\
{\tt $^6$ ESA, P.O. Box 78, 28691 Villanueva de la Ca\~nada, Madrid, Spain}\\
{\tt $^7$ ESO, Karl Schwarzschild Strasse 2, D85748, Garching, Germany}\\
{\tt $^8$ Instituto de Astrofisica de Canarias, V'a Lactea, E38205, La Laguna (Tenerife), Spain}\\
{\tt $^9$ Dipartimento di Astronomia, Universita' di Bologna, Via Ranzani 1, 40127, Bologna, Italy}\\
{\tt $^{10}$ INAF-Osservatorio Astronomico di Bologna, Via Ranzani 1, 40127, Bologna, Italy}} \maketitle
\begin{abstract}
This paper presents the first direct estimate of the 3D clustering properties of far-infrared sources 
up to $z\sim 3$. This has been possible thanks to the Pacs Evolutionary Probe (PEP) survey of the GOODS South field performed with the PACS instrument 
onboard the{ \it Herschel} Satellite. 
550 and 502 sources were detected respectively in the 100$\mu$m and in the 160$\mu$m channels down to fluxes $S_{100\mu \rm m} = 2$ mJy and 
$S_{160\mu \rm m} = 5$ mJy, cuts which ensure $>$80\% completeness of the two catalogues. More than 65\% of these sources has an 
(either photometric or spectroscopic) redshift determination from the MUSIC catalogue, this percentage rising to $\sim 95$\% in the inner portion 
of GOODS South which is covered by data at other wavelengths.  
An analysis of the de-projected two-point correlation function $w(\theta)$ over the whole redshift range spanned by the data reports for the 
(comoving) correlation length, $r_0\sim 6.3$ Mpc  and $r_0\sim 6.7$ Mpc, respectively at 100$\mu$m and 160$\mu$m, corresponding to dark matter halo 
masses $M\simgt10^{12.4} M_\odot$, in excellent agreement with previous estimates obtained for mid-IR selected sources in the same field.
Objects at $z\sim 2$ instead seem to be more strongly clustered, with  $r_0\sim 19$ Mpc and $r_0\sim 17$ Mpc in the two considered PACS channels.
This dramatic increase of the correlation length between $z\sim 1$ and $z\sim 2$ is connected 
with the presence, more visible at 100$\mu$m than in the other band, of a wide (at least 4 Mpc across in projection), $M\simgt 10^{14} M_\odot$, 
 filamentary structure which includes more than 50\% of the sources detected at $z\sim 2$. 
An investigation of the properties of such sources indicates the possibility for boosted star-forming activity in those which reside within the 
overdense environment with respect of more isolated galaxies found in the same redshift range. If confirmed by larger datasets, this result 
can be explained as due to the combined effect of large reservoirs of gas available at high redshifts in deep potential wells such as those associated 
to  large overdensities and the enhanced rate of encounters between sources favoured by their relative proximity. 
Lastly, we also present our results on the evolution of the relationship between luminous and dark matter in star-forming galaxies between 
$z\sim 1$ and $z\sim 2$. We find that the increase of (average) stellar mass in galaxies $<M_*>$  between $z\sim 1$ and $z\sim 2$ is about a factor 10 
lower than that of the dark matter haloes hosting such objects 
($<M_*>^{z\sim 1}/<M_*>^{z\sim 2} \sim 4 \cdot 10^{-1}$ vs $M_{\rm halo}^{z\sim 1}/M_{\rm halo}^{z\sim 2} \sim 4 \cdot 10^{-2}$). 
When compared with recent results taken from the literature, 
our findings agree with the evolutionary picture of downsizing wherby massive galaxies at $z\sim 2$ were more actively forming stars than 
their $z\sim 1$ counterparts, while at the same time contained a lower fraction of their mass in the form of luminous matter. 
\end{abstract}
\section{Introduction}
Investigations of the Large Scale Structure (LSS) traced by selected classes of extra-galactic sources, read in connection with more and more refined 
models for its characterization, has enabled in the recent years to derive important information on some physical properties characterizing the objects 
producing the clustering signal. One of these key quantities is the halo mass and its evolution with look back time, which for instance allows to derive 
cosmological connections between different populations of extra-galactic sources observed in different wavelengths at different epochs. 

The development of these theoretical models for the description of the LSS, together with the impressive achievements reached in the past fifteen years 
or so by science instrumentation have allowed the scientific community to greatly broaden their knowledge on the formation and evolution of the cosmic 
structures, pushing it up to the dawn of galaxy formation. 

Within this framework, amongst the more interesting classes of sources are those which are undergoing some kind of active process such as AGN activity or star formation.
Since the first high redshift observations of the two Hubble Deep Fields (Williams et al. 1996; Casertano et al. 2000), all clustering analyses performed on optical/near-IR surveys which were at the same time wide enough to ensure statistical significance and deep enough to probe high redshift regimes, have highlighted a strong evolution of these two classes of sources, whereby activity shifts to smaller and smaller haloes/galaxies as one moves forward in time. The critical redshift for this transition seems to be around $z\sim 2$ for star-forming galaxies (e.g. Magliocchetti \& Maddox 1999; Arnouts et al. 2002; Foucaud et al. 2010; Hartley et al. 2010) and $z\sim 3$ for QSOs (e.g. Porciani, Magliocchetti \& Norberg 2004; Shen et al. 2007; Ross et al. 2009). Perhaps not surprisingly, these two redshift values bracket the peak of cosmic activity both in terms of AGN and star formation (Madau et al. 1996).  

Although clustering results in flux-limited surveys are somehow masked by the well known 'Malmquist bias' issue (more distant and therefore more luminous sources are more strongly clustered), some steps forward have been taken in order to minimize its effect on high redshift studies (e.g. Magliocchetti et al. 2008; Foucaud et al. 2010). These works all converge at indicating that massive galaxies form at high redshifts and on short timescales, while the sites of activity shift to lower mass
systems at later epochs. This pattern, referred to as "downsizing" (Cowie et al. 1996; Heavens et al. 2004), is also observed in an increasing number of independent works 
(see e.g. Fontana et al. 2004; Treu et al. 2005; Cimatti, Daddi \& Renzini 2006, Saracco et al. 2006; Bundy et al. 2006; McLure et al. 2006 just to mention a few), all adopting techniques which differ from clustering analyses.

The advent of the {\it Spitzer} telescope has marked another milestone for what concerns our knowledge of the population of dusty (and therefore optically dim) active galaxies. 
Even in this case, the downsizing effect was confirmed for all obscured active sources up to $z\sim 2.5$ (e.g. Magliocchetti et al. 2008).
In particular, for the first time Spitzer has detected a population of  Ultra-Luminous Infrared Galaxies (ULIRGS, $L\simgt 10^{12} L_\odot$) in the redshift range ($1.6\simlt z\simlt 2.5$) with extremely high clustering lengths ($r_0\sim 15$ Mpc -- Farrah et al. 2006b; Magliocchetti et al. 2007; 2008; Brodwin et al. 2008). Taken at face value, these correlation lengths correspond locally to those of groups-to-poor clusters of galaxies (see e.g. Guzzo et al. 2000; Estrada et al. 2009), and might be interpreted with a roughly one-to-one correspondence between the overwhelming majority of ULIRGS at $z\sim 2$ and local clusters.

The recent advent of the {\it Herschel} telescope allows the scientific community to further extend  its knowledge on the formation and evolution of dusty sources up to very high redshifts. 
In fact, thanks to the fantastic capabilities of the PACS instrument, the GOODS South region has been observed at 70$\mu$m, 100$\mu$m and 160$\mu$m to unprecedented depths as a part of the PEP (PACS Evolutionary Probe, D. Lutz et al. submitted to A\&A) survey. The  majority of PEP sources in the GOODS-S is endowed with a redshift determination from the MUSIC catalogue (Santini et al. 2009) and this allowed us to estimate, for the first time in a self-consistent way, the 3D clustering properties of the population of far-infrared galaxies up to $z\sim 3$. Knowledge of the redshift distribution $N(z)$ further allows to investigate the evolution of clustering with look-back time. In the light of what was discussed above, we have then also investigated the large-scale pattern traced by PEP galaxies at redshifts $z\sim 2$ and 
will discuss our findings in terms of abundances of super-structures or proto-clusters made of obscured star forming galaxies right  at the peak of cosmic star formation activity.

The work is organized as follows: \S 2 introduces the PEP catalogues and analyses some of the properties of the sources  with the help of multiwavelength data coming from the MUSIC catalogue. \S\S 3 and 4 present the results for the two-point angular correlation function of all PEP sources spanning the whole $z=[0-3]$ redshift range. \S 5 concentrates on the large-scale properties of far-infrared sources residing in the $z\sim 2$ epoch which corresponds to the peak of cosmic star formation activity. \S 6 provides a more detailed description of $z\sim 2$ PEP sources, in particular focussing on the properties of an extended filamentary structure (or proto-cluster) made of 100$\mu$m-selected galaxies at $z\sim 2.2$.  \S 7 presents our results on the 
evolution of the relationship between luminous and dark matter in PEP-selected galaxies and compares them with recent works from the literature, 
while \S 8 summarizes our findings. 
  
Throughout this paper we assume a $\Lambda$CDM cosmology with $H_0=70 \: \rm Km\:s^{-1} 
 Mpc^{-1}$, $\Omega_0=0.27$ and $\Omega_\Lambda=0.73$. 


\section{The data}

\begin{figure*}
\begin{center}
\vspace{8cm}  
\includegraphics{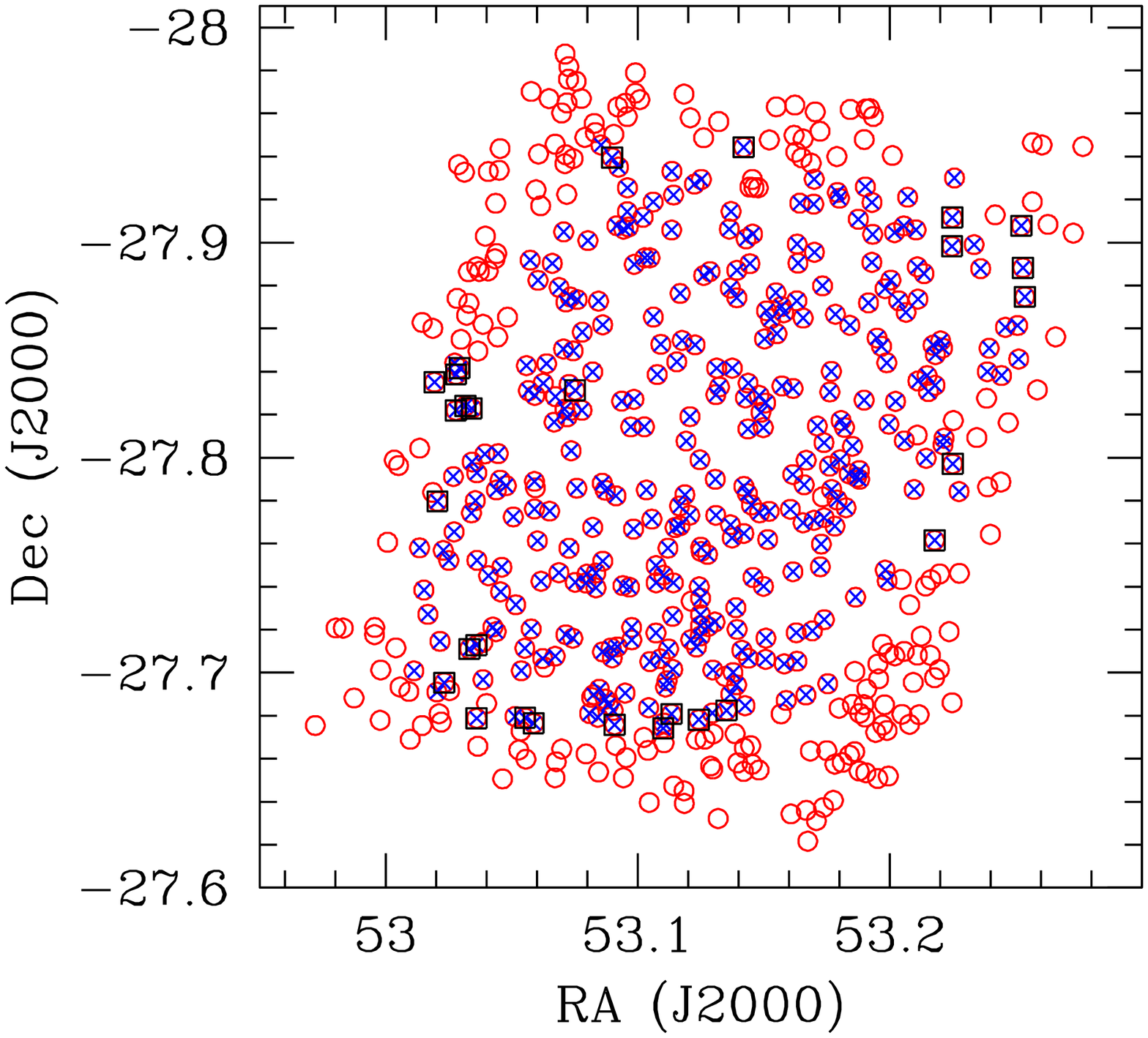} \includegraphics{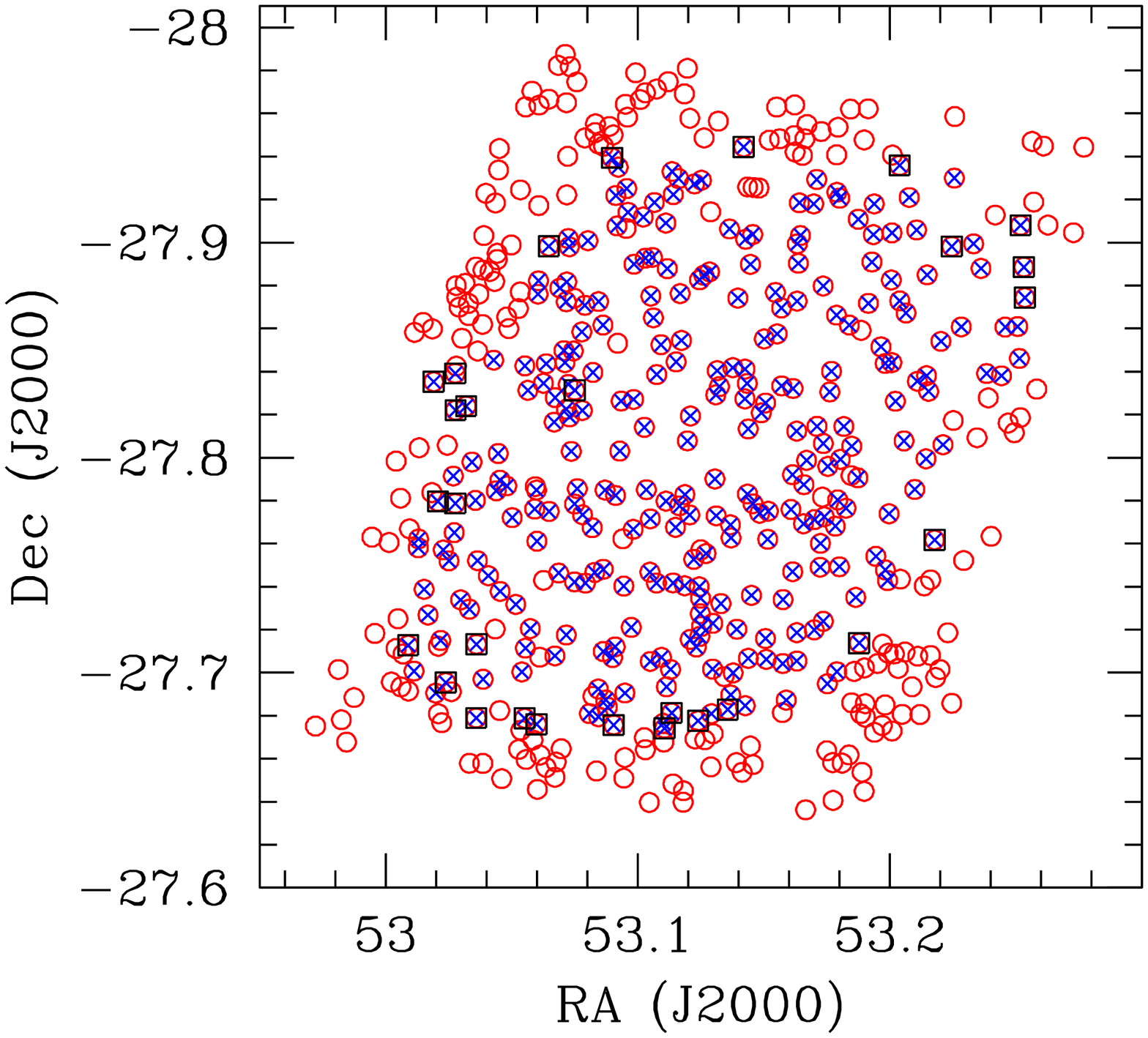}
\caption{Projected distribution of sources in the GOODS-S  down to $\sim 80$\% completeness limit at 100$\mu$m (left-hand panel) and 160$\mu$m (right-hand panel). Crosses indicate those sources which have a counterpart  in the MUSIC catalogue (Santini et al. 2009), while the squares highlight objects which, although present in the MUSIC catalogue, lack of a redshift determination (either photometric or spectroscopic -- see text for details). 
\label{fig:goodss}}
\end{center}
\end{figure*}

\begin{figure*}
\begin{center}
\vspace{8cm}  
\includegraphics{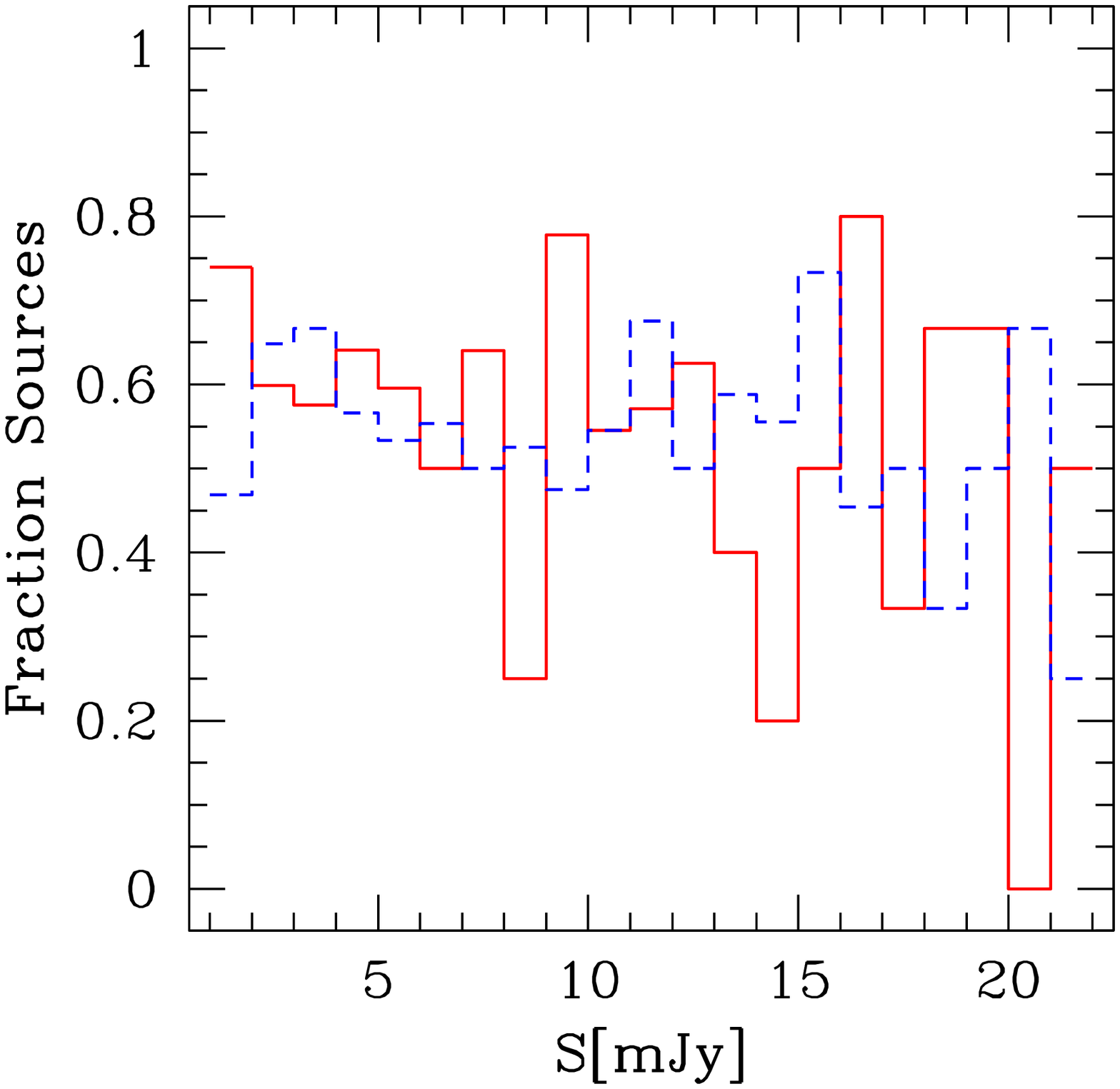} \includegraphics{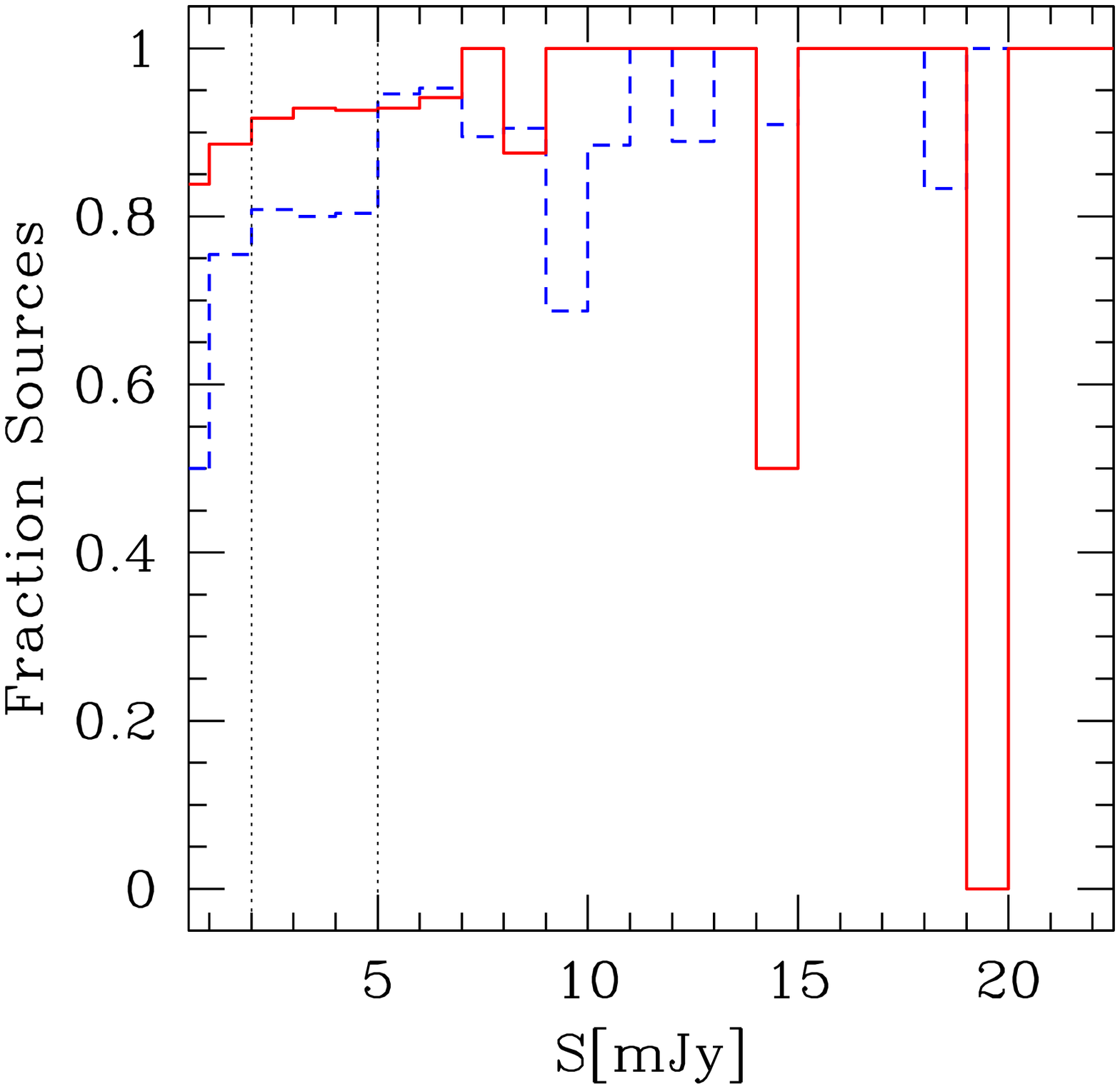}
\caption{Fraction of sources with either a spectroscopic or a photometric redshift determination in GOODS-S as a function of far-IR flux. The solid line represents PEP 100$\mu$m observations, while the dashed one is derived at 160$\mu$m.  The left-hand panel refers to the whole area surveyed by PEP, while the right-hand one only concentrates on those PEP sources falling in the inner region of GOODS South which is also covered by MUSIC data.
\label{fig:flux_ratio}}
\end{center}
\end{figure*}

The GOODS South (GOODS-S) region has been observed by the PACS (Poglitsch et al. 2010) instrument onboard the {\it Herschel} Space Observatory (Pilbratt et al. 2010) as a part of the PACS Evolutionary Probe (PEP,  D. Lutz et al. submitted to A\&A) Survey, aimed at  studying the properties and cosmological evolution of the infrared population up to redshifts $z\sim 3-4$. We refer to the Lutz et al. paper for further information on the survey and fields, including GOODS South, data reduction and catalogue creation.

The total number of sources detected in the blind (i.e. without the use of priors) catalogues at the  $\simgt 3\sigma$ confidence level is  375 at $70\mu$m, 717 at 100$\mu$m and 867 at 160$\mu$m. The corresponding flux limits in the three wave-bands are $\sim 1.2$ mJy, $\sim 1.2$ mJy and $\sim 2.0$ mJy. Given the relative paucity of sources in the blue/70$\mu$m channel which does not allow sensible clustering estimates, this paper will only concentrate on the two higher wavelengths, i.e. the $100\mu$m and 160$\mu$m bands. 

GOODS-S is one of the best studied cosmological fields, having been observed in great depth at virtually all the available wavelengths. The GOODS-MUSIC catalogue (Santini et al. 2009) provides a compilation of high quality multi-wavelength (from 0.3$\mu$m to 24$\mu$m) data for sources detected in the GOODS-S area in the z-, Ks-, and 4.5 $\mu$m-bands. The largest fraction of the sample is 90\% complete at z$\sim$26 or Ks$\sim$23.8 or 4.5um$\sim$23.5 (AB scale; Oke, 1970). The final catalogue comprises 15,208 sources, out of which nearly 3,000 are endowed with a spectroscopic redshift determination. Most of the remaining sources have assigned photometric redshifts, estimated by following the technique of Grazian et al. (2006). 

The original $\ge 3\sigma$ PEP catalogues at the two relevant frequencies were then cross-correlated with the MUSIC catalogue through a three-band maximum likelihood procedure  (Sutherland \& Saunders, 1992; see Berta et al. 2010 for further details). This found 
respectively 475 (out of 717) and  536 (out of 867) IR-selected galaxies with a counterpart in MUSIC. The overwhelming majority of these objects has an assigned (either photometric or spectroscopic) redshift. In a handful of cases this was not possible due to the lack of mainly JHK band data at the edges of the area covered by MUSIC. We have also checked 
the accuracy of the photometric redshift estimates for our population of
 star-forming galaxies. The mean spread is found to be 
$\left<(z_{\rm spec}-z_{\rm phot})/(1+z_{\rm spec})\right>\simeq 0.076$, only slightly 
larger than the value found by Santini et al. (2009) for the whole 
GOODS-MUSIC catalogue. By investigating this issue further, we found that such an 
increased value is mainly due to a handful (about 6) of outliers with either $z_{\rm spec}$ 
 or $z_{\rm phot}$ beyond $\sim 2$. Indeed, if one removes these few sources, 
the mean spread in photometric accuracy is about 0.049, i.e. even smaller than that 
reported by Santini et al. (2009).\\
\\

\noindent
Clustering studies need high completeness levels. This is why in the following analysis we will only concentrate on those sources belonging to the blind PEP catalogues which have PACS fluxes above the $\sim$80\% completeness levels. These limits, as estimated from simulations (cfr. Lutz et al. submitted to A\&A) are 2 mJy at $100\mu$m and 5 mJy at 160$\mu$m.  
Applying the above flux cuts, the number of sources respectively in the green and red bands of the blind catalogues are 550 and 502. These are the sources which will be used for the clustering studies. Their projected distribution onto the sky is shown in Figure \ref{fig:goodss} by the empty circles.

Of these sources, respectively 343 at 100$\mu$m and 296 at 160$\mu$m have a counterpart in the MUSIC catalogue, obtained as previously explained. Their distribution onto the sky is represented in Figure \ref{fig:goodss} by the crosses. From the two plots it is possible to appreciate that almost all PEP sources in the inner GOODS-S area have a counterpart in the MUSIC catalogue, the apparent discrepancy between the total number of PEP-detected objects and of those which also have a counterpart in other wavebands simply due to the fact that the PEP survey covers a slightly wider area than those at other frequencies.\\
 The number of PEP-MUSIC sources above the 80\% completeness level with an assigned redshift is 315 and 283, respectively at 100$\mu$m and 160$\mu$m. Those sources which do have a counterpart in MUSIC but for which it was not possible to assign a redshift estimate are indicated in Figure \ref{fig:goodss} by the empty squares.
As already discussed, they mostly lie at the edges of the area covered by MUSIC data which lack of JHK band data coverage.

\begin{figure}
\begin{center}
\vspace{8cm}  
\includegraphics{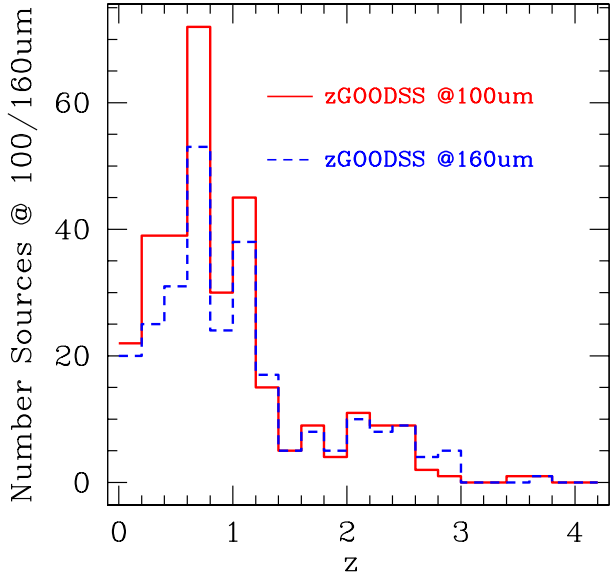} 
\caption{Redshift distribution for the $>80$\% complete samples of PACS-selected sources in the GOODS-S. The solid line refers to the 100$\mu$m sample, while the dashed line is at 160$\mu$m. 
\label{fig:Nz}}
\end{center}
\end{figure}

The ratio between the total number of PEP sources and of  those with a redshift determination as a function of PACS fluxes  is presented in Figure \ref{fig:flux_ratio} by the histograms, solid line for the 100$\mu$m case, dashed one at 160$\mu$m. The left-hand panel refers to all the sources detected by the PEP survey, while the right-hand one only considers those PEP-selected objects which fall within the area covered by MUSIC data. As it is possible to notice, the success rate for redshift determination is roughly constant with far-IR flux, at least for fluxes above the 80\% completeness level (represented in the right-hand panel by the dotted vertical lines). Furthermore, 
the right-hand panel also indicates that in the common region covered by both MUSIC and PEP observations, more than 90\% of the PEP sources above the chosen flux limits are endowed with a redshift determination. We will discuss the implications of these results in \S 3.2 when dealing with conversions to spatial estimates of the angular correlation function.

The redshift distribution for those PEP-selected sources with a counterpart in the MUSIC catalogue and with PACS fluxes above the 
$80$\% completeness level is presented in Figure \ref{fig:Nz}. The distributions at the two frequencies are virtually identical, with a first peak around $z\sim 0.7$, a secondary peak at $z\sim 1.1$, then a rapid decrease before presenting a wider bump between redshifts 2 and 3. The two primary peaks correspond to the two well studied clusters at $z=0.73$ and $z=1.1$ (e.g. Gilli et al. 2003; Vanzella et al. 2005).

\section{Clustering Analysis} 

\begin{figure*}
\begin{center}
\vspace{8cm}  
\includegraphics{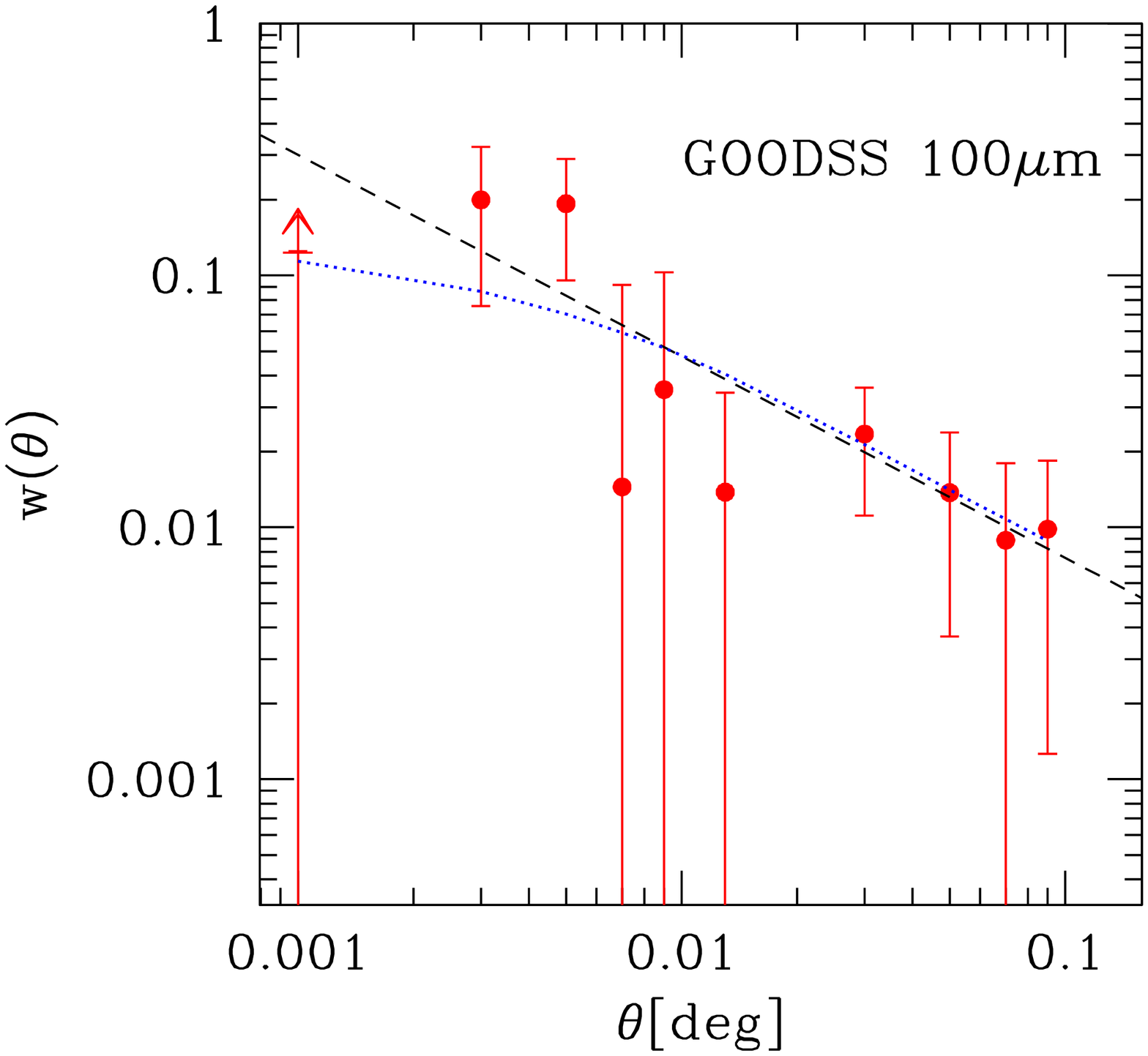} \includegraphics{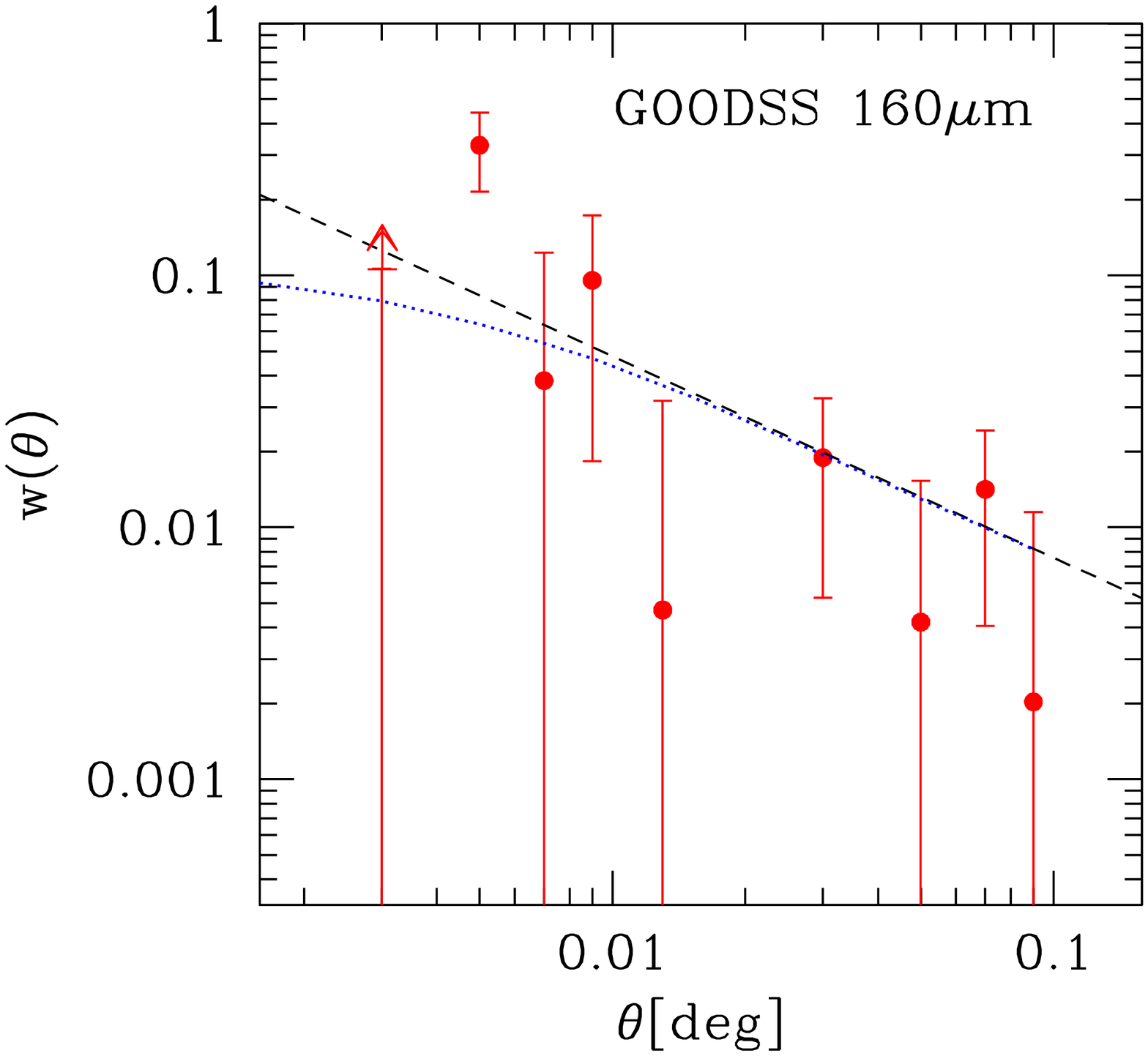}
\caption{Angular correlation function for PACS-selected sources above the 80\% completeness level in the GOODS-S at 100$\mu$m and 160$\mu$m. The dashed lines indicate the best power-law fits to the data according to equation  (\ref{eqn:limber}), with  
$r_0=6.3$ Mpc and $r_0=6.7$ Mpc, respectively at 100$\mu$m and 160$\mu$m, while the dotted lines indicate the best-fits to the data according to equation (\ref{eq:bias}) with $M_{\rm min}=10^{12.4} M_{\odot}$ at both sampled wavelengths.  Upward arrows correspond to the lower limits in the estimates of $w(\theta)$ in the lowest angular scale intervals due to instrument resolution (see text for details).
\label{fig:w_theta}}
\end{center}
\end{figure*}

\subsection{The Angular Correlation Function}
The angular two-point correlation function
$w(\theta)$ gives the excess probability, with respect to a
random Poisson distribution, of finding two sources in the solid
angles $\delta\Omega_1$ $\delta\Omega_2$ separated by an angle
$\theta$. In practice, $w(\theta)$ is obtained by comparing the
observed source distribution with a catalogue of randomly
distributed objects subject to the same mask constraints as the
real data. \\
We chose to use the estimator (Hamilton 1993)
\begin{eqnarray}
w(\theta) = 4\times \frac{DD\cdot RR}{(DR)^2} -1, 
\label{eq:xiest}
\end{eqnarray}
where $DD$, $RR$ and $DR$ are the number of data-data, random-random 
and data-random pairs separated by a distance $\theta$. 

Given the $>80$\% completeness of both the 100$\mu$m and 160$\mu$m considered samples, 
in the specific case of  PACS-selected sources we have simply generated random catalogues with twenty times as many objects as the real data sets, covering the whole surveyed GOODS South area minus the outermost regions which showed irregular data coverage. Within the same considered area, the number of $S_{100\mu \rm m}\ge 2$ mJy and $S_{160 \mu \rm m}\ge 5$ mJy  PEP-selected sources is respectively 519 and 470. 
$w(\theta)$ in eq. (\ref{eq:xiest}) for the above objects was then estimated in the angular range 
$10^{-3}\simlt \theta \simlt 0.1$ degrees. The upper limit was determined by 
the geometry of the GOODS-S field, and corresponds to about half of the maximum scale 
probed by the survey. 

Figure 4 presents our results for the two 100$\mu$m-selected  (4a; left-hand panel) 
and 160$\mu$m-selected (4b; right-hand panel) samples. 
The error bars show Poisson estimates for the points. Since the distribution is clustered, these estimates only provide a lower limit to the uncertainties. However, it can be shown  
that over the considered range of angular scales this estimate is close to 
those obtained from bootstrap resampling (e.g. Willumsen, Freudling 
\& Da Costa, 1997).  Due to the angular resolution of the Herschel mirror at the two considered 
wavelengths (respectively $\sim 8^{\prime\prime}$ and $\sim 13^{\prime\prime}$ at 100$\mu$m and 160$\mu$m) which would cause source blending on lower angular scales, 
the two results in the smallest $\theta$ bins of the angular correlation function  merely represent lower limits and therefore have been indicated with upward arrows.

If we then assume the standard power-law form for $w(\theta)=A\theta^{1-\gamma}$,
we can estimate the parameters $A$ and $\gamma$ by using a least-squares
fit to the data.  Given the large errors on $w$,  
we chose to fix $\gamma$ to the standard value $\gamma=1.8$.
Although somewhat arbitrary, this figure and its assumed lack of dependence 
on redshift is justified by Large Scale Structure observations of large enough samples of 
high redshift sources so to allow for a direct estimate of the slope of $w(\theta)$
at different look back times (e.g. Porciani, Magliocchetti \& Norberg 2004; 
Le Fevre et al. 2005).
The small area of the GOODS-S survey introduces a negative bias through
the integral constraint, $\int w^{est} d\Omega = 0 $. We allow for
this by fitting to $ A \theta^{1-\gamma} -C $, where $C = 1.78 A$.\\
The dashed lines in Figures~4a and 4b represent the best power-law fits 
respectively to the 100$\mu$m and 160$\mu$m data-sets. 
The associated best-fit values for the amplitude are 
$A^{100\mu \rm m}=[1.2\pm 0.4]\cdot 10^{-3}$ for the 100$\mu$m-selected sample, 
and $A^{160\mu \rm m}=[1.2\pm 0.5]\cdot 10^{-3}$ for the 
160$\mu$m-selected one, i.e. the results at the different wavelengths are virtually identical.

As a further consistency check, we have released the constraint on $\gamma$ and 
let it vary between 1.0 and 2.5. Despite the expected degeneracy 
between the best-fit values of $A$ and $\gamma$ determined by this choice, 
we find that in the case of the 100$\mu$m-selected sample 
slopes shallower than 1.3 are ruled out by our data at the $>95$\% confidence level. 
$\gamma=1.8$ (and the corresponding value for $A$ quoted above) indeed provides the 
best fit to the observed $w(\theta)$, even though any value in the range 
$1.3\le\gamma\le 2.2$ can be accepted within 1$\sigma$ confidence level. 
The corresponding ranges of variability 
 for the amplitude are $A^{100\mu \rm m}\sim[0.2-19.0]\cdot 10^{-3}$.
The 160$\mu$m-selected sample (for which, however, $w(\theta)$ is less constrained due to a worse angular resolution of 
the corresponding PACS channel) instead seems to prefer higher values of $\gamma$. This is 
due to the observed $w(\theta)$ at the smallest angular scales probed by our analysis 
and possibly hints at 
multiple occupancy for the dark matter haloes inhabited by such galaxies (cfr \S 4). 
The best-fit value is found for $\gamma\ge 2.5$ (outside the chosen range of variability for $\gamma$), 
with a 1$\sigma$-acceptance that goes down to $\gamma=1.9$. The corresponding ranges of variability 
for the amplitude are $A^{160\mu \rm m}\sim[(\le 0.6) - 8]\cdot 10^{-4}$.\\

The estimated amplitudes for these samples (under the assumption of $\gamma=1.8$)
are in good agreement 
with the results obtained for Spitzer-selected sources. 
More specifically,  they closely follow those obtained by Magliocchetti et al. (2007) 
for the whole population of $F_{24\mu \rm m} \ge 0.35$~mJy objects
from the First Look Survey ($A=[9\pm 2]\cdot 10^{-4}$), and also those obtained by Fang et al. (2004) for IRAC, 8$\mu$m-selected galaxies in the same field. Lastly, we also note that the low clustering signal obtained by Maddox et al. (2010) for galaxies selected at 250$\mu$m  within the Herschel-ATLAS consortium is compatible with our PEP results.

\subsection{Relation to spatial quantities}
The standard way of relating the angular two-point correlation
function $w(\theta)$ to the spatial two-point correlation function
$\xi(r,z)$ is by means of the relativistic Limber equation (Peebles,
1980):
\begin{eqnarray}
w(\theta)=2\:\frac{\int_0^{\infty}\int_0^{\infty}F^{-2}(x)x^4\Phi^2(x)
\xi(r,z)dx\:du}{\left[\int_0^{\infty}F^{-1}(x)x^2\Phi(x)dx\right]^2},
\label{eqn:limber} 
\end {eqnarray}
where $x$ is the comoving coordinate, $F(x)$ gives the correction for
curvature, and the selection function $\Phi(x)$ satisfies the relation
\begin{eqnarray}
{\cal N}=\int_0^{\infty}\Phi(x) F^{-1}(x)x^2 dx=\frac{1}{\Omega_s}
\int_0^{\infty
}N(z)dz,
\label{eqn:Ndense} 
\end{eqnarray}
in which $\cal N$ is the mean surface density on a surface of solid angle
$\Omega_s$ and $N(z)$ is the number of objects in the given survey
within the shell ($z,z+dz$). 

If we adopt a spatial correlation function 
of the form $\xi(r,z)=(r/r_0)^{-1.8}$, independent of redshift
 in the considered intervals, and we consider the 
redshift distributions presented in Figure~\ref{fig:Nz}, 
for the adopted cosmology we obtain via a $\chi$-square fit to the data: $r_0^{100\mu \rm m}=6.3^{+1.1}_{-1.3}$~Mpc and 
$r_0^{160 \mu \rm m}=6.7^{+1.5}_{-1.7}$~Mpc (where both quantities are comoving), respectively for the $100\mu$m and $160\mu$m samples. 

It is worth noting that, while $w(\theta)$ has been obtained 
over the whole area covered by PACS observations of the GOODS South, the adopted redshift distributions only refer to a sub-sample of $\sim$65\% of the PEP sources which are located in the (smaller) region where PEP and MUSIC data co-exist (cfr. Figure \ref{fig:goodss}). 
In principle, such a discrepancy could produce biases in the determination of $r_0$. However,
as seen in \S 2, the ratio between the number of PEP sources endowed with a redshift determination and that of the whole PEP sample shows a general independence of far-IR flux. Furthermore, this trend is the same both in the inner and outer regions of the PEP-GOODSS survey (cfr. Figure \ref{fig:flux_ratio}). 
These findings provide insurance on the reliability of the adopted $N(z)$ and therefore on the goodness of the estimated $r_0$. 

As a further consistency check, we have also considered the MUSYC catalogue of photometric redshifts  derived by Cardamone et al. (2010) over the whole Extended Chandra Deep Field-South region and assigned via a nearest neighbourhood procedure a photometric counterpart to our PEP sources detected in the blind catalogues. By then adopting the redshift distributions originating from this different approach, we find for the correlation lengths $r_0^{100\mu \rm m}\simeq 6.2$~Mpc and  $r_0^{160\mu \rm m}\simeq 6.7$~Mpc, i.e. the results are virtually equivalent to those obtained by considering the redshift distributions derived from the MUSIC catalogue.

The above results indicate that the clustering properties of far-infrared galaxies closely follow those found for 24$\mu$m-selected sources. Gilli et al. (2007), while analysing a sample of 
$S_{24\mu\rm m}\ge 20\mu$Jy in the same GOODS South field we are considering in this work, 
report a very similar clustering length  ($r_0\sim 7.3$ Mpc) as ours for their sub-class of 
Luminous Infrared Galaxies (LIRG) with $L>10^{11}L_\odot$. 
Noticeably, also the number densities of their 24$\mu$m-selected sources and of our far-infrared galaxies coincide, leading us to conclude that these two classes of objects refer to the same galaxy population. It is also worth noticing that Gilli et al. (2007) find a dip in the trend of 
the real-space correlation function on scales $\sim 1$ Mpc which, at the median redshift ($z\sim 1$) of their survey, corresponds to those angular scales where our data also seem to indicate lower-than-expected values for $w(\theta)$.
The above clustering properties for far-infrared bright galaxies locally mirror 
those of 'normal' early-type galaxies (e.g. Madgwick et al. 2003; Zehavi et al. 2005) and it is therefore sensible to envisage an evolutionary connection between these populations whereby the bulk of far-IR selected sources at $z\sim 1$ will eventually evolve in passive ellipticals in the nearby universe.   

All the information on the adopted samples, correlation function results, number densities etc. can be found summarized in Table 1.

\section{Connection with physical properties: the Halo Bias Approach}

The easiest way to relate the clustering trend of a population of extra-galactic sources with their physical properties is by means of the Halo Bias approach (Mo \& White 1996) 
which describes the observed clustering signal as the product between the two-point correlation function of the underlying dark matter distribution $\xi_ {\it dm}(r,z)$ and the square of the so-called 
bias function $b(M_{\rm min},z)$,  quantity which at a given redshift solely depends on the minimum mass $M_{\rm min}$ of the haloes in which the detected astrophysical objects reside via the relation: 
\begin{equation}
\xi(r,z)=\xi_ {\it dm}(r,z) \cdot b(M_{\rm min},z)^2.
\label{eq:bias}
\end{equation}
Starting from eq.(\ref{eq:bias}), the theoretical angular two-point correlation function $w(\theta)_{\rm th}$ predicted by the halo bias model 
 for PACS-selected galaxies at 100$\mu$m and 160$\mu$m has then been computed once again by following eq.(\ref{eqn:limber})  and the 
best values for $M_{\rm min}$ at the two wavelengths have been found by a $\chi^2$-fit to the 
observed clustering signal. This gives ${\rm Log}M_{\rm min}=12.4^{+0.2}_{-0.5}M_\odot$ and 
${\rm Log}M_{\rm min}=12.4^{+0.4}_{-0.6}M_\odot$, respectively for the 100$\mu$m and 160$\mu$m samples. The corresponding best-fit curves are shown in Figures 4a and 4b by the (blue) dotted lines. 

More sophisticated models have been introduced in the recent years in order to provide a more realistic description of the observed clustering properties of extra-galactic sources.
One of the most successful ones, the so-called Halo Occupation Model (HOM, e.g. Scoccimarro et al. 2001) relates the clustering properties of a chosen population of galaxies with the way such objects populate their dark matter haloes, i.e. relaxes the somehow unrealistic assumption, implicit in the halo bias model,  of having a one-to-one correspondence between a galaxy and its dark matter halo.  
Indeed, from a quick glance at Figures 4a and 4b, it is possible to notice how the more simple halo bias model (\ref{eq:bias}) cannot reproduce the observed rise in the angular correlation function on small angular scales ($\theta \simlt 0.05$ deg),  where the signal is dominated by pairs of galaxies residing within the same dark matter halo.  
However desirable,  such an analysis is unfortunately not possible on our dataset due to relatively poor signal-to-noise ratio. In fact, the HOM is a three parameter model: together with the minimum mass of the dark matter halo it also depends on the the quantities $\alpha$ and $N_0$ according to the expression 
\begin{eqnarray}
N(M) = N_0(M/M_{\rm min})^{\alpha} \;\;\;
\rm{if}\ M \ge  M_{\rm min}, \nonumber
\label{eq:Ngal}
\end{eqnarray}
where $N(M)$ is the number of galaxies within a halo of certain mass $M$.  And a three-parameter fit will only be possible when larger fields such as COSMOS and the Lockman Hole 
will be included in the analysis.
For the time being, we notice that the quantities $M_{\rm min}$ and $\alpha$ are covariant, so that within the HOM framework, higher values for $\alpha$ in general correspond to lower 
values for $M_{\rm min}$. 
This implies that in the presence of multiple halo occupancy, the values for $M_{\rm min}$ 
found within the HOM scenario will be (slightly) lower than those obtained via the halo bias model adopted in this work. 

Despite the above warnings and bearing in mind that so far all clustering results at the SPIRE wavelengths have only been derived by using theoretical redshift distributions, it is however interesting to note that the resulting values for $M_{\rm min}$ 
are in good agreement with those found by Cooray et al. (2010) in their 250$\mu$m-selected sample obtained by combining the Lockman-SWIRE and FLS fields, possibly confirming the fact that the imaged sources belong to the same underlying population of galaxies as ours. We also note that similar results have also been obtained by Magliocchetti et al. (2008) for $0.6\simlt z\simlt 1.2$ galaxies with $S_{24\mu\rm m}\ge 400\mu$Jy taken from the UKIDSS survey, once again indicating that the bulk of sources producing the clustering signal at the PACS wavelengths coincides with $z\sim 1$ Spitzer-selected galaxies.


\section{Large-Scale Structure at $z\sim 2$}

\begin{figure*}
\begin{center}
\vspace{8cm}  
\includegraphics{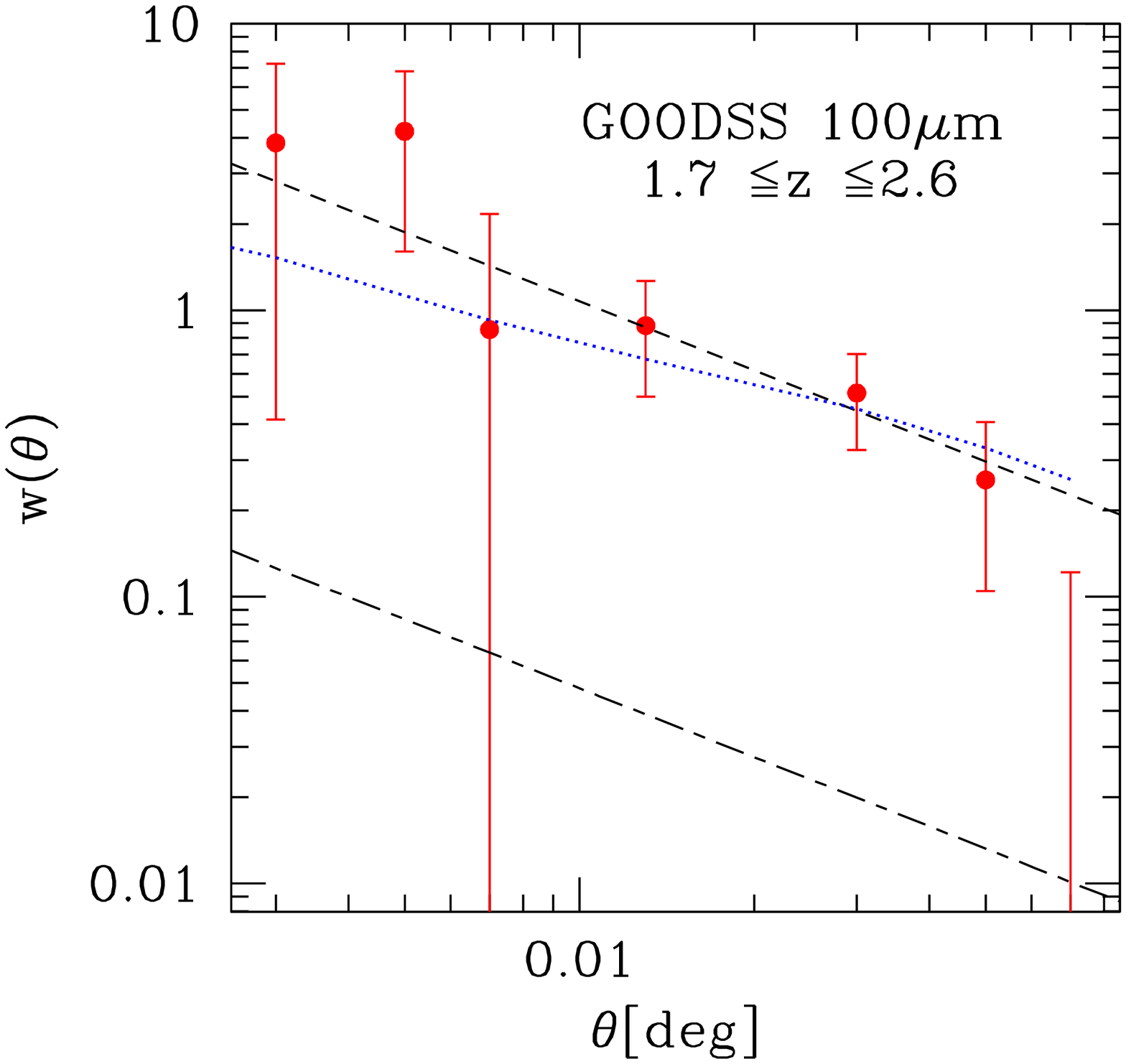} \includegraphics{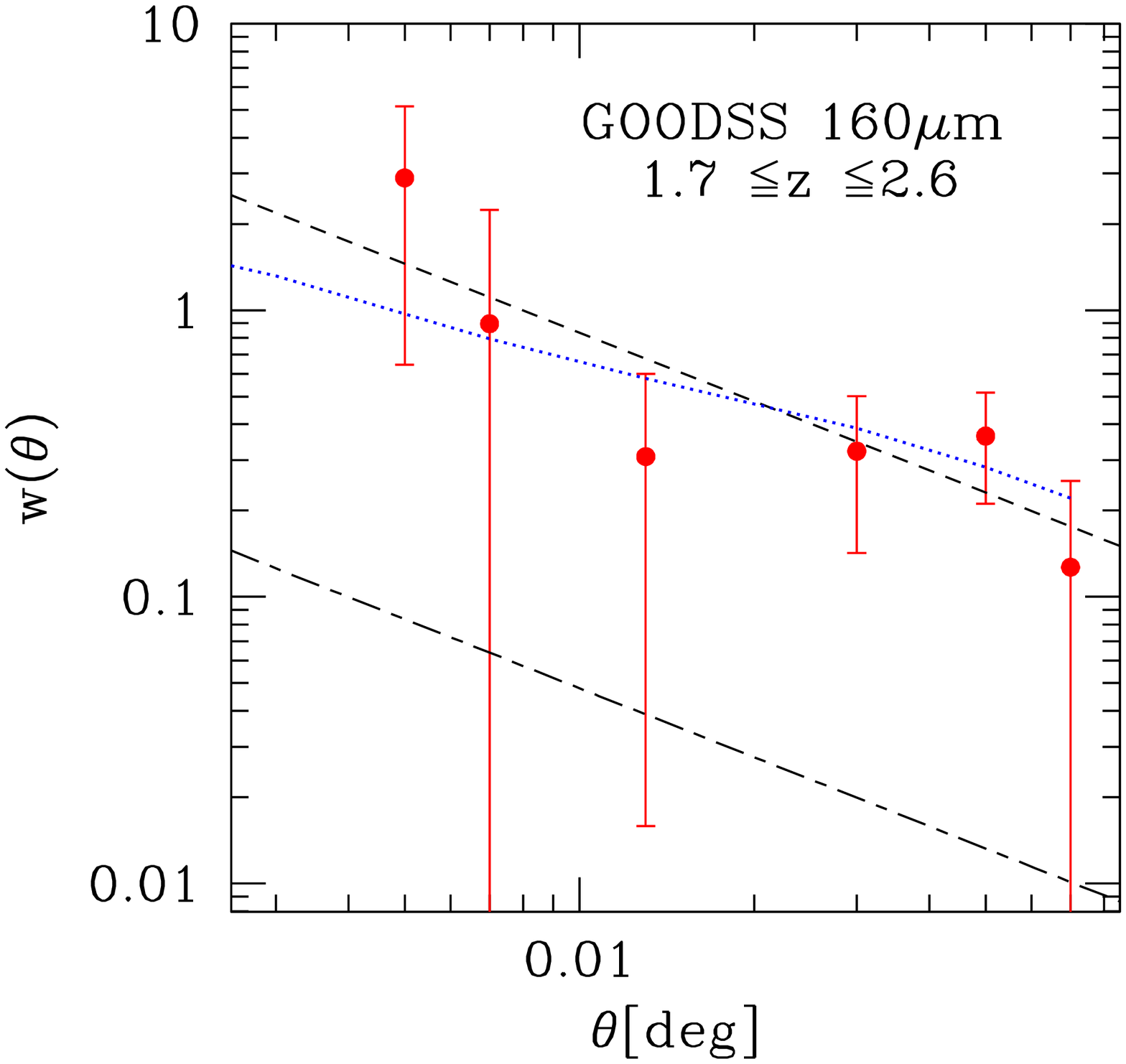}
\caption{Angular correlation function of PACS-selected,  $z=[1.7-2.6]$, sources with fluxes above the 80\% completeness level in the GOODS-S at 100$\mu$m (left-hand panel) and 160$\mu$m (right-hand panel). The dashed lines indicate the best power-law fits to the data obtained for   
$r_0=19.0$ Mpc and $r_0=17.4$ Mpc, respectively at 100$\mu$m and 160$\mu$m, while the dotted lines indicate the best-fit to the data according to equation (\ref{eq:bias}) with $M_{\rm min}=10^{13.8} M_{\odot}$ (left-hand panel) and $M_{\rm min}=10^{13.7} M_{\odot}$ (right-hand panel). The long-short dashed lines in both panels represent the best fits to the data presented in Figure 4. 
\label{fig:w_highz}}
\end{center}
\end{figure*}

\begin{figure*}
\begin{center}
\vspace{8cm}  
\includegraphics{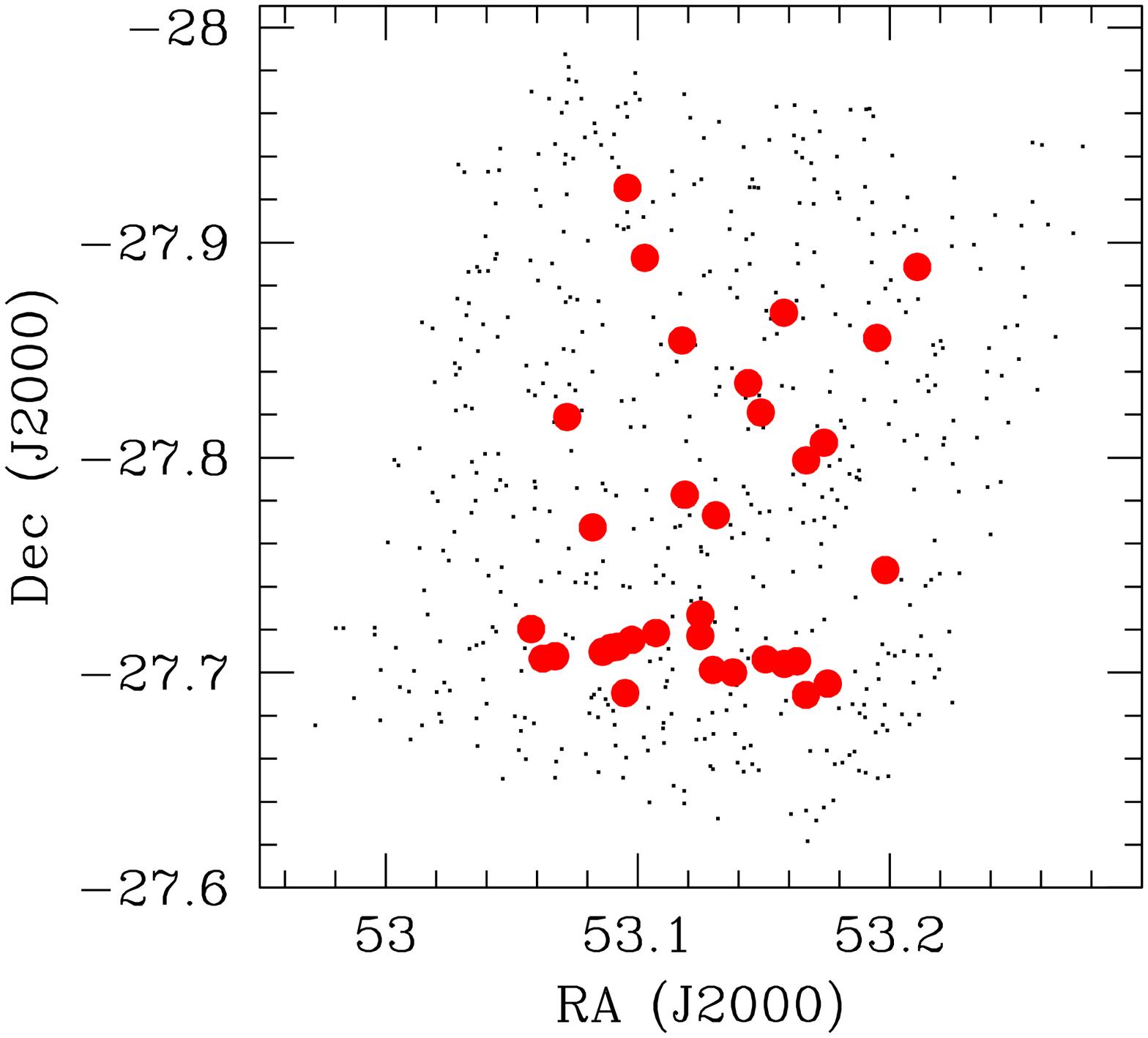} \includegraphics{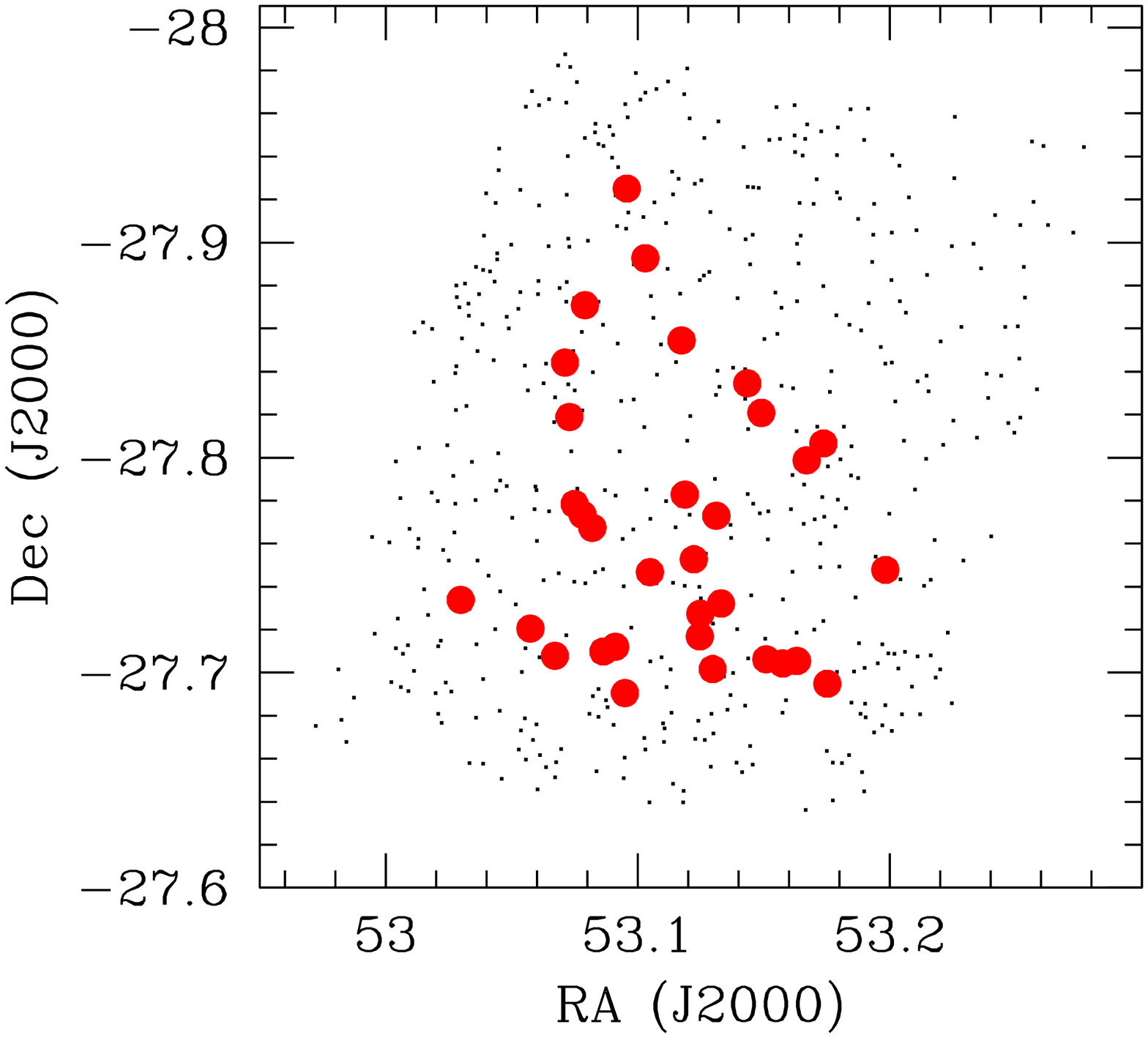}
\caption{Distribution of $z=[1.7-2.6]$ sources in the GOODS-S with fluxes above $80$\% completeness limit (100$\mu$m on the left-hand side, 160$\mu$m on the right-hand side). The small dots represent the whole PEP population brighter than the same flux cuts.
\label{fig:goodss_highz}}
\end{center}
\end{figure*}

\begin{table*}
\begin{center}
\caption{Properties of sources in the blind GOODS-S at 100$\mu$m and 160$\mu$m: a) total number
above 80\% completeness level, in brackets those used for the clustering analyses; b) amplitude $A$ of the angular two-point correlation function $w(\theta)$; c) comoving correlation length $r_0$; d) minimum dark halo mass $M_{\rm min}$; number density}
\begin{tabular}{cccccc} 
         & Number sources & $A$ & $r_0$ [Mpc] & Log$M_{\rm min} $ [$M_\odot$]& number density $[\rm Mpc^{-3}]$\\
\hline
\hline
100$\mu$m; all $z$& 550 (519) &$(1.2\pm 0.4) \times 10^{-3}$&  $6.3^{+1.1 }_{-1.3} $&$12.4^{+0.2 }_{-0.5}$ & $(2.0\pm 0.1)\times 10^{-3}$\\
100$\mu$m; $1.7 \le z  \le 2.6$& 33 (33) &$(2.7\pm 0.7) \times 10^{-2}$&  $19.0^{+2.6 }_{-2.9} $&$13.8^{+0.2 }_{-0.3}$ & $(7.4\pm 1.3)\times 10^{-5}$\\
\hline
160$\mu$m; all $z$& 502 (470) &$(1.2\pm 0.5) \times 10^{-3}$&  $6.7^{+1.5 }_{-1.7} $&$12.4^{+0.4 }_{-0.6}$ & $(1.7\pm 0.1)\times 10^{-3}$\\
160$\mu$m; $1.7 \le z  \le 2.6$& 32 (32) &$(2.0\pm 0.6) \times 10^{-2}$&  $17.4^{+2.8 }_{-3.1} $&$13.7^{+0.3 }_{-0.4}$ & $(7.2\pm 1.3)\times 10^{-5}$\\
\hline
\hline

\end{tabular}
\end{center}
\end{table*}

Magliocchetti et al. (2007; 2008) found that Spitzer/24$\mu$m-selected galaxies residing at redshifts $z\sim 2$ exhibited an exceptionally strong clustering signal 
($r_0\sim 16$ Mpc). Their results were also found to agree with the estimates of Farrah et al. (2006a; 2006b) for ultra-luminous infrared galaxies over the $1.5\simlt z \simlt 3$ range. 
The above figure, obtained for galaxies undergoing an intense phase of dust-enshrouded star formation (in the redshift interval $1.7\simlt z \simlt 2.6$, the strong spectral PAH feature centred at 7.7$\mu$m -- indicative of a very intense star-forming activity -- falls in the 24$\mu$m band),
was also in agreement with earlier results obtained for star-forming galaxies selected at $z\sim 2$ in the (rest-frame) UV-band in both Hubble Deep Fields (Magliocchetti \& Maddox 1999; Arnouts et al. 2002). Although with larger uncertainties, sub-millimeter-selected galaxies -- mainly found in the considered redshift range -- also seem to cluster with a similar strength (Weiss et al. 2009). More recently, Hartley et al. (2010) use the UKIDSS Ultra Deep survey to investigate the clustering properties of passive and star-forming galaxies and find that, while locally these two classes of objects exhibit quite a different behaviour, with the former one being about twice as more clustered than the second one, at redshifts $\sim 2$ star-forming and passively evolving galaxies 
have similar clustering strengths, with values for the correlation length in excellent agreement with those reported above ($r_0\sim  14$ Mpc). 
We note that at lower redshifts, such a strong clustering signal finds a counterpart only in the case of  radio sources (see e.g. Magliocchetti et al. 2004) and early type, $L\simgt 4\;L^*$ 
galaxies (see e.g. Norberg et al. 2002),  and is only second to those of rich
clusters of galaxies (e.g. Guzzo et al. 2000). It is therefore natural to 
envisage a connection between these latter objects and the high-z sources, 
whereby massive galaxies undergoing intense star-formation at redshifts 
$z\simgt 2$ end up as the very bright central galaxies 
(passive objects with a high 
probability for enhanced radio activity, see e.g. Best et al. 2007; 
Magliocchetti \& Br\"uggen 2007) of local clusters.

Redshifts around 2 therefore seem very interesting not only for what concerns the study of the assembling of the stellar content of galaxies (it is in fact at those redshifts that the cosmological star formation rate peaks --  Madau et al. 1996, see also Gruppioni et al. 2010 for a more updated version of the Madau plot), but also with respect to the formation and evolution of extremely large structures such as groups and clusters of galaxies. 
It is for this reason that we have decided to isolate in the 100$\mu$m and 160$\mu$m PEP catalogues sources in the redshift range $1.7\le z\le 2.6$ -- so to mimic the 24$\mu$m selection 
of galaxies with intense star forming activity and therefore strong PAH lines -- and have then studied their Large-Scale properties. 
The number of sources in the PEP/GOODS South catalogues within the above redshift interval and with fluxes above the requested cuts which ensure $>80$\% completeness is 33 in the 100$\mu$m case and one less at 160$\mu$m. 
The angular correlation function of these objects has then been calculated as in \S 3.1.

The corresponding results with associated (Poissonian) uncertainties are plotted in Figure \ref{fig:w_highz}, where once again those referring to the 100$\mu$m band are found on the left-hand panel. The short-long dashed lines indicate the best-fit to the angular correlation functions obtained for the whole PEP samples, both at 100$\mu$m and at 160$\mu$m (Figure 4 and Table 1). Despite the large uncertainties, comparisons with these latter results clearly show that $z\sim 2$ PEP galaxies are extremely strongly clustered, with an amplitude of about a factor 20 higher than what found for PEP sources at all redshifts. 
A more quantitative analysis performed by assuming a power-law form for $w=A \theta^{(1-\gamma)}$, with $\gamma=1.8$ (see \S 3), returns the values for the amplitude $A^{100\mu \rm m}=[2.7\pm 0.7]\cdot 10^{-2}$ for the 100$\mu$m-selected sample, 
and $A^{160\mu \rm m}=[2.0\pm 0.6]\cdot 10^{-2}$ for the 
160$\mu$m-selected one. In terms of correlation lengths (cfr. eq. \ref{eqn:limber}), this corresponds to $r_0^{100\mu \rm m}=19.0^{+2.6}_{-2.9}$~Mpc and 
$r_0^{160 \mu \rm m}=17.4^{+2.8}_{-3.1}$~Mpc. These values (which we stress are obtained by adopting a top-hat distribution for $N(z)$ which does not include the (small, cfr \S2) errors 
associated to the photometric redshift estimates)
 are extremely high, possibly even higher than those obtained in the works summarized at the beginning of this section in the case of star-forming galaxies at redshifts $z\sim 2$. We also note that according to the halo bias approach (equation \ref{eq:bias}), such very high clustering strengths imply $z\sim 2$ PEP sources to be hosted in dark matter haloes of masses respectively greater than $13.8^{+0.2 }_{-0.3}M_\odot$  and $13.7^{+0.3 }_{-0.4} M_\odot$ at 100$\mu$m and 160$\mu$m, i.e. these sources reside in super-massive/cluster-like structures which are already found in place at $z\sim 2$. 
It is also interesting to stress that the above values agree with the early estimates of Daddi et al. (2001),   $r_0\sim 12-24$ Mpc, obtained for a handful of spectroscopically determined $z\sim 2$ GOODS-S galaxies.

A visual investigation of the distribution of PEP-selected galaxies in the $z=[1.7-2.6]$ interval 
(cfr. Figure \ref{fig:goodss_highz}) shows that this is indeed the case. Out of 33 sources with $S_{100 \mu\rm m}\ge 2$ mJy, 18 of them, corresponding to $\sim$55\% of the total, reside in a huge, filamentary structure,  about 4 Mpc across in projection, centred at $z\simeq 2.2$. A similar behaviour, although less pronounced (which explains the slightly lower value obtained for the correlation length in this latter case) is also visible at 160$\mu$m.  We note that a forming, diffuse, super-structure at $z\sim 2.2$ was already identified by Salimbeni et al. (2009), while studying the properties of overdensities in the GOODS South. It is however remarkable the way this overdensity shows up in such a clean way and with such a high significance in a numerically much smaller sample which is uniquely made of dusty star-forming galaxies.

We know however that the GOODS South seems to be a very peculiar field, already dominated by a couple of clusters around redshifts 0.7 and 1 (e.g. Gilli et al. 2003; Adami et al. 2005;
Vanzella et al. 2005; D'az-Sanchez et al. 2007). So, it is sensible to state the statistical significance of our findings. Straight integration of the Sheth \& Tormen (1999) dark halo mass function in the considered $z=[1.7-2.6]$ redshift range 
indicates that the number of haloes with masses greater than $10^{13.5} M_\odot$, compatible with our findings, on an area like that subtended by the GOODS-South is about 14. This corresponds to the 15, 100$\mu$m-selected, real 'field' (i.e. not belonging to the filamentary super-structure) galaxies observed with Herschel-PACS with $S_{100 \mu\rm m}\ge 2$ mJy. From the above figures we can then draw a first, interesting, conclusion:  {\it every dark matter halo which reaches the mass necessary to host a (dusty) star-forming galaxy at redshift $z\sim 2$ will be inhabited by (at least) one of such objects}. In other words, 
the above statement indicates that star-forming galaxies at redshifts around 2 are common events, 
their observed paucity (number densities of the order of a few $10^{-5}$Mpc, cfr. Table 1) simply due to the paucity of high mass dark matter haloes at the considered redshifts. In a similar fashion, we can also conclude that, for what concerns the spatial distribution of $z\sim 2$ dust-obscured galaxies, GOODS South is not a particularly peculiar field as it comprises (within Poissonian uncertainties) a number of sources which is compatible with theoretical expectations.

Similar back-of-the-envelope calculations further show that the number density associated to the projected distribution of $S_{100 \mu\rm m}\ge 2$ mJy proto-cluster sources is about $2\times 10^{-4}$Mpc, with a $\delta_{\rm protocluster}/\delta_{\rm field} \simeq 5$, i.e. as already expected, the overdensity at $z\sim 2.2$ , is not yet virialized  ($\delta_{\rm vir} \sim 178$ for a $\Lambda$CDM universe; Peebles 1980) but will likely evolve in a bound structure.


\section{The $z\sim 2.2$ overdensity}

\begin{figure}
\begin{center}
\vspace{8cm}  
\includegraphics{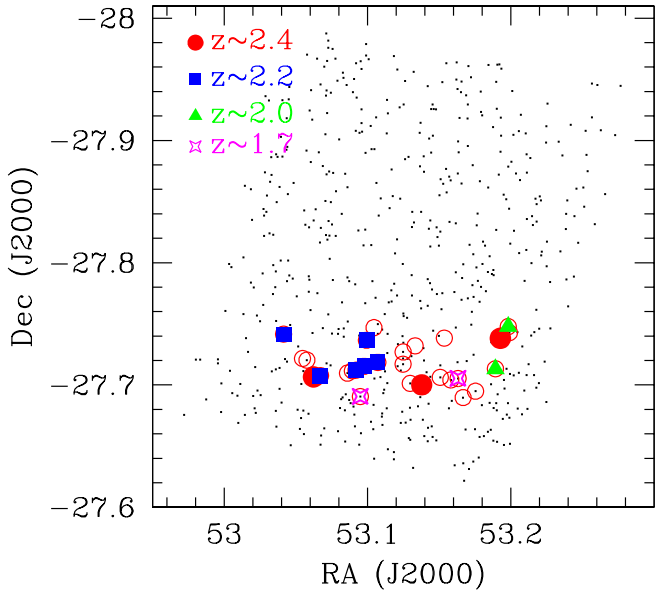} 
\caption{Distribution of all PEP/100$\mu$m-selected sources belonging to the $z\sim 2.2$ overdensity  down to $S_{100 \mu\rm m}\simeq 1$ mJy. Empty circles are for objects with estimated photometric redshifts from MUSIC, while different symbols indicate sources with measured spectroscopic redshifts in various z slices around the $z=2.2$ value.
\label{fig:cluster}}
\end{center}
\end{figure}

The $z\sim 2.2$ filamentary structure singled out in the high redshift clustering analyses has an undoubtable relevance {\it per se} as it provides one of the first clear examples of formation of super-structures made of star-forming galaxies right at the peak of cosmic star-forming activity (for a further example cfr. Andreani, Magliocchetti \& de Zotti 2010).  It is therefore worth concentrating some more on this structure by analyzing its members and their properties as compared to those of similar galaxies residing in less dense environments.

In order to do that, we have released the request for $>$80\% completeness and considered all PEP/100$\mu$m-selected sources within the $1.7\le z\le 2.6$ redshift interval which in projection appear to belong to this diffuse overdensity. The number of objects fulfilling the above requirements is 28 and they are indicated in Figure \ref{fig:cluster} with (red) empty circles.
Thirteen of these galaxies have a spectroscopic redshift determination from the literature and are highlighted in Figure \ref{fig:cluster} with different symbols. More in detail, three of them, distributed along the whole length of the filamentary structure, have $z\sim 2.4$ (filled red dots), six of them have $z\sim 2.1-2.2$ and are all clustered in the left-hand part of the structure (filled blue squares), while respectively two+two sources have redshifts around 2.0 and 1.7, the former pair residing in the outermost right-hand part of the overdensity (green triangles), while the other, probably not really belonging to the diffuse super-structure,  is spread along the proto-cluster (magenta stars). It therefore seems that we are in the presence of a very diffused/possibly forming super-structure at $z\sim 2.4$ 'bracketed' at the two ends of its length by two more concentrated clumps set at a median redshift $z\sim 2.1$ (with a scatter $\Delta z\sim 0.1-0.2$).

\begin{figure*}
\begin{center}
\vspace{8cm}  
\includegraphics{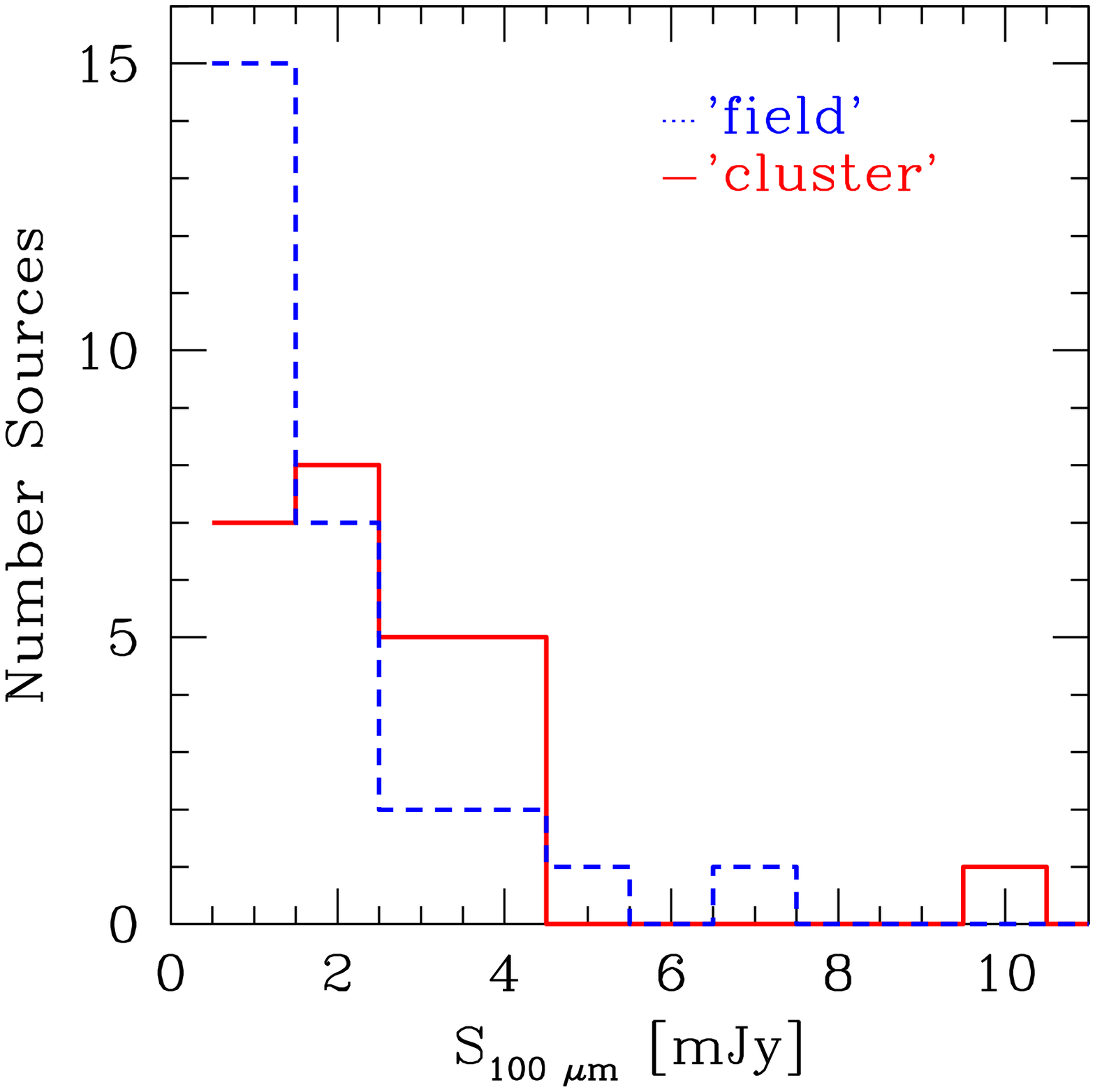} \includegraphics{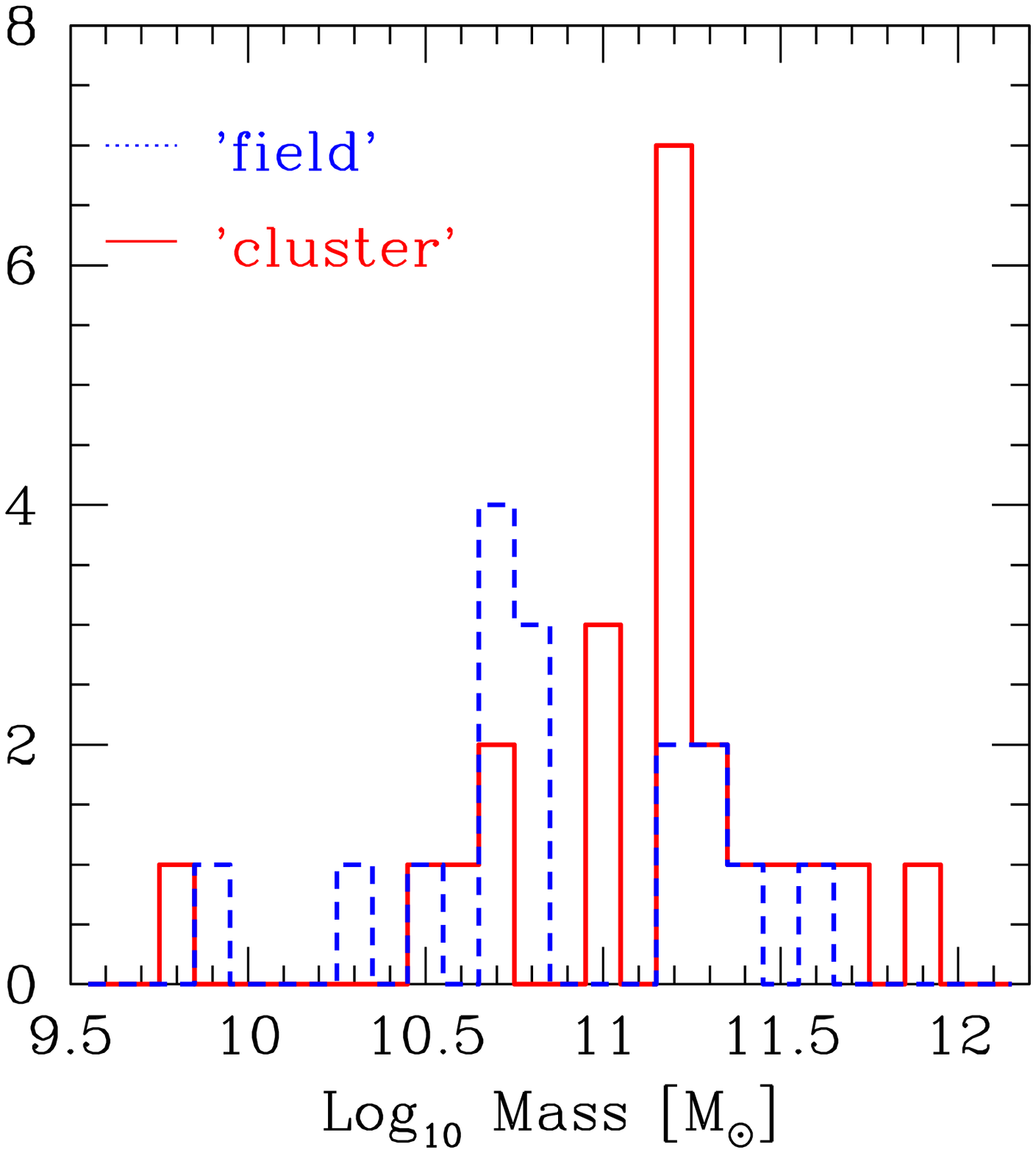}\includegraphics{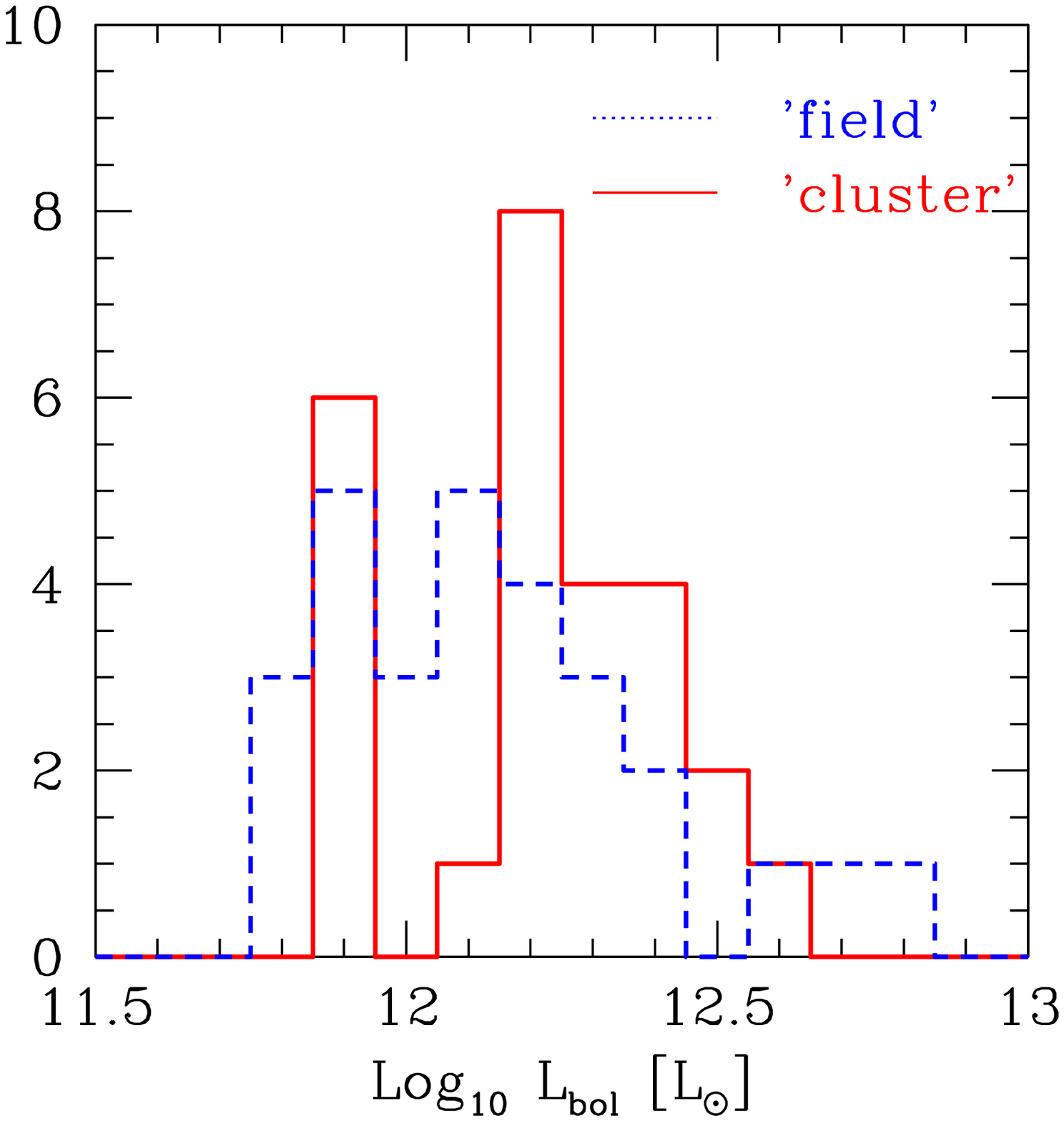}
\caption{Distribution of 100$\mu$m fluxes (left-hand panel), masses (middle panel) and bolometric luminosities (right-hand panel) for PEP/100$\mu$m-selected galaxies in the redshift range $1.9\simlt z\simlt 2.6$. The solid line represents galaxies belonging to the proto-cluster structure, the dotted line the others ('field' galaxies).
\label{fig:hist_lum}}
\end{center}
\end{figure*}


In order to investigate the influence of environment on star-forming activity at redshifts $\sim 2$, we have then estimated masses and bolometric luminosities for the 25 $1.9\simlt z\simlt 2.6$ sources residing within the area identified by the super-structure and then compared these latter quantities with those obtained for more isolated or 'field' galaxies (28 objects), otherwise selected from the 100$\mu$m-PEP sample with the same criteria as the sources belonging to the overdensity.

Stellar masses have been derived according to the method described in Rodighiero et al. (2007; 2010) which fits the UV-to-5.8$\mu$m broad band photometric data with the Bruzual \& Charlot (2003) stellar population synthesis models, assuming a Salpeter (Salpeter 1955) Initial Mass Function (IMF)  and a Calzetti extinction law (Calzetti et al. 1994), once the redshift of each source is fixed to its spectroscopic or, whenever unavailable, photometric value.  For GOODS-S the stellar masses are also available from Santini et al. (2009) and are calculated with the same 
synthesis models and very similar recipes as in Rodighiero et al. (2007; 2010).

Bolometric luminosities $L_{\rm bol}$ have instead been computed by using two different methods. 
The first one, presented in Rodighiero et al. (2010), fits the optical-to-FIR SEDs with a library of local templates, including those of Polletta et al. (2007), and a few additional modified templates with colder FIR emission (see Gruppioni et al. 2010 for further details) at a given redshift. 
$L_{\rm bol}$  was then obtained by integrating the best-fit SEDs over the whole $[8-1000]\mu$m rest-frame range. The second method instead follows that presented in Santini et al. (2009).
Luminosities in this case have been derived by fitting the 100$\mu$m observations to the Dale \&
Helou (2002) synthetic templates. First, a total infrared luminosity in the $[8-1000]\mu$m range was assigned to each template by using the empirical relation of Marcillac et al. (2006)
between $L_{\rm bol}$ and the predicted ratio between fluxes at 60$\mu$m and 100$\mu$m. Once the k-correction was applied, we then selected the model which best reproduces
the observed 100$\mu$m luminosity  and normalized it by using the flux
difference between the model and the observed galaxy. These two methods produce virtually identical results, so in the following we only present the outcomes of the Santini et al. (2009) method.

Figure \ref{fig:hist_lum} presents the distribution of 100$\mu$m fluxes (left-hand panel), masses (middle panel) and luminosities (right-hand panel) for the galaxies belonging to the filamentary $z\sim 2.2$ overdensity (solid line) as compared to those of more isolated galaxies residing within the same redshift range (dotted line). As it is possible to appreciate, while the distribution of masses in the two cases follows a  similar trend, 'field' galaxies seem to be on average fainter than sources belonging to the overdensity, both in terms of 100$\mu$m flux and bolometric luminosity. If we then convert bolometric luminosities into estimates for the star formation rate (SFR) via the relation $\rm SFR [\rm M_\odot/yr]=1.8\times 10^{-10} L_{\rm bol}/L_\odot$ (Kennicutt et al. 1998), it follows that Figure \ref{fig:hist_lum} also provides a snapshot for the 
star formation activity within and outside the enhanced-density area. 
By doing this, and despite the relative small statistics, we find that the sources within the filamentary structure have an enhanced star formation activity (peak at $\sim 300 M_\odot\rm /yr$) with respect to more isolated galaxies, which exhibit a peak in their SFR at about 180$M_\odot/\rm yr$.  We are therefore in the presence of boosted star-formation activity in the over-dense region, probably due to the joint effect of large reservoirs of gas available at high redshifts in deep potential wells such as those associated to large overdensities and the enhanced rate of encounters between sources favoured by their proximity. Larger datasets are obviously needed to provide the ultimate answer on this issue. 
This result is in agreement with the findings of Popesso et al. (2010) who observe an inversion of the relationship between environmental density and star formation rate at around redshift 1 with respect to the local results, where SFR is depressed in overdense regions.



\section{On the redshift evolution of the link between luminous and dark matter}
Since GOODS-South has been observed at a great wealth of frequencies, important information such as mass in stars, bolometric luminosity 
and star formation rate for the sources which populate the area can be obtained. A combination of these properties with the results inferred from the 
clustering analysis on the mass of their dark matter haloes can then provide interesting constraints on the relation between -- say -- luminous and dark 
matter in star forming galaxies and its eventual dependence on look-back time. Given the extreme similarity in the clustering results obtained in this 
work between the 100$\mu$m and 160$\mu$m channels (cfr. Table 1), and the better determination of the 
relevant parameters derived in the former case due to its higher angular resolution, in the following we will only concentrate on galaxies selected at 
100$\mu$m.

Stellar masses and star formation rates were then derived for all sources detected by PEP at 100$\mu$m in the GOODS-South field endowed with a reliable 
(either photometric or spectroscopic) redshift determination. This was done as described in \S 6. Furthermore, in order to ensure both completeness of 
the sample and consistency with the clustering results, we have restricted our analysis to galaxies with 100$\mu$m fluxes brighter than 2 mJy. Finally, 
in order to investigate possible redshift dependencies of the considered quantities, we have considered two different samples: a first one which 
comprises galaxies in the $0.7 \le z \le 1.2$ range ($N_{\rm gal}=82$ with both stellar mass and luminosity estimates), and a second one which focuses 
on sources at higher, $1.7\le  z\le 2.6$, redshift ($N_{\rm gal}=32$ with both stellar mass and luminosity estimates). 

Future works will address in more detail the properties
of the sources detected by the PEP survey in GOODS-S. We here discuss average stellar
masses and star formation rates of these two samples in conjunction with the minimum dark
matter halo mass $M_{\rm min}$ as determined from clustering results.\\
The mean redshifts of the two considered samples are $<z>=0.94$ and $<z>=2.19$. 
Average stellar masses, bolometric luminosities and star formation rates are found to be  $<M_*> = 10^{10.6\pm 0.4} M_\odot$, 
$<L_{\rm bol}>= 10^{11.4\pm 0.3} L_\odot$ and $<SFR>= 52\pm 35\; M_\odot/\rm yr$   for the intermediate-redshift sample and
 $<M_*> = 10^{11\pm 0.3} M_\odot$, $<L_{\rm bol}>= 10^{12.3\pm 0.2} L_\odot$ and $<SFR>= 444\pm 210 \;M_\odot/\rm yr$ for the high-z one. 

Despite the uncertainties due to the relatively small number of sources considered in the analysis, the above figures suggest that by moving towards 
the present universe not only
 -- as expected for a flux-limited survey --
we are selecting less luminous and actively star-forming galaxies, but also that 
on average such galaxies have smaller stellar masses than their high-redshift counterparts. 
As we saw in \S\S 4 and 5, this is also true when we consider the masses of the dark matter halos which host such objects. However, the increase 
of mass with redshift in this latter case seems to be more 
pronounced. Indeed, while for stellar masses we find a ratio $<M_*(<z>=0.94)>/<M_*(<z>=2.19)>\sim 0.4$, for the minimum mass of dark matter haloes 
we obtain $M_{\rm min}(<z>\sim 1)/M_{\rm min}(<z>=2.19)\sim 0.04$ (cfr. Table 1), i.e. a value which is about ten times smaller. \\
We note that, although for the calculation of $M_{\rm min}(<z>\sim 1)$ we have technically used the whole sample of 100$\mu$m-selected galaxies brighter 
than 2 mJy, the presence of the two very marked peaks in their redshift distribution at $z\sim 0.7$ and $z\sim 1.1$ (cfr. Figure 3) effectively 
ensures that the observed clustering signal is dominated by sources which reside at redshifts which are close to those of the two peaks and therefore 
that the measured values for $M_{\rm min}$ closely resemble those that would be obtained for a redshift-limited, $0.7\simlt z\simlt 1.2$, sample such 
as that considered in this section.

A number of recent works have focussed on both theoretical (e.g. Moster et al. 2010) and observational (e.g. Foucaud et al. 2010; Leauthaud et al. 2011; 
Wake et al. 2011) aspects of the relationship between stellar and total mass in galaxies as a function of both redshift and galaxy masses. 
The observational works deal with clustering analyses and their results can therefore be compared with what we obtain in the present paper. 
In more detail, both Foucaud et al. (2010) and Wake et al. (2011), despite the different selection criteria (galaxies selected in the near-IR rather 
than in the far-IR regime as it is our case), span a redshift and a stellar mass range which roughly overlap with ours. Foucaud et al. (2010) find for 
galaxies with stellar masses respectively within the values $10^{10.5}M_\odot\le M_* \le 10^{11.0}M_\odot $ and $10^{11}M_\odot\le M_* \le 10^{12}M_\odot $ 
at $<z>=1$ and $<z>=1.73$ values for the number density $n_g$ and clustering length $r_0$, 
$n_g(<z>=1)=(9\pm 1)\cdot 10^{-4}$ Mpc$^{-3}$, $r_0(<z>= 1)=8.0\pm 1.4$ Mpc  and $n_g(<z>=1.7)=(6 \pm 2)\cdot 10^{-5}$ Mpc$^{-3}$, $r_0(<z>=1.7)=22\pm 9$ 
Mpc, in very good agreement with our findings. 
The same can be said of the Wake et al. (2011) results who find in their $<z>=1.1$ and $<z>=1.9$ redshift bins respectively 
$n_g(<z>=1.1)\sim 2 \cdot 10^{-3}$ Mpc$^{-3}$, $r_0(<z>=1.1)=8.0\pm 0.5$ Mpc and $n_g(<z>=1.9)\sim 8.5 \cdot 10^{-5}$ Mpc$^{-3}$, 
$r_0(<z>=1.9)=16 \pm 2$ Mpc for stellar masses $M_*/M_\odot(<z>=1)\ge 7 \cdot 10^9$ and $M_*/M_\odot(<z>=1.9)\ge10^{11}$. 


The concordance between the above results and ours also implies that our findings on the relationship between luminous and dark matter, i.e. the ratio 
$M_*/M_{\rm min}$ at redshifts $\sim 1$ and $\sim 2$ agree with what reported by both Foucaud et al. (2010) and Wake et al. (2011). We also note that, 
even though direct comparisons are more difficult given the way data are presented, at $z\sim 1$ our findings are also 
in agreement with the results from  Leauthaud et al. (2011) who add galaxy lensing to their analysis, and with 
those of Meneux et al. (2008). 

Although the intrinsic limitations of dealing with a flux-limited catalogue mask possible strong conclusions that can be drawn on the evolution 
of the luminous-to-dark matter ratio in galaxies with redshift/galaxy mass, however the pieces of information that we have 
gathered in this section, when combined with the above results taken from the literature, suggest that at higher redshifts PEP-selected galaxies 
were more actively forming stars than their lower-z counterparts. At the same time, they also contained a lower fraction of their mass in (evolved) 
stars. These findings then converge at indicating an evolutionary picture whereby the bulk of stellar formation within massive galaxies took place 
at $z\simgt 2$ (in excellent agreement with the results on the cosmic evolution of the star formation rate -- e.g. Madau et al. 1996). 
The intensity of this dramatic process then decreases as we approach the local universe, and already by $z\sim 1$ star formation activity 
within massive galaxies is largely reduced as most of the stars within such galaxies have already been formed.


\section {conclusions}
We have presented  the first direct estimates of the 3D correlation properties of far-infrared-selected sources up to redshifts $z\sim 3$. This has been possible thanks to the Pacs Evolutionary Probe (PEP) survey of the GOODS South field performed with the PACS instrument onboard the 
{\it Herschel} Satellite. 550 and 502 sources were detected respectively in the 100$\mu$m and in the 160$\mu$m channels down to fluxes $S_{100\mu \rm m} = 2$ mJy and $S_{160\mu \rm m} = 5$ mJy, cuts which ensure $>$80\% completeness of the two catalogues. 

These two samples have then been cross-correlated with the MUSIC (Santini et al. 2009)
catalogue in order to provide counterparts at other wavelenghts to the PEP sources and, where possible, to also assign them a (either photometric or where available spectroscopic) redshift. This was possible in more than 65\% of the cases, independent of infrared flux. This apparent incompleteness in the success rate for redshift determination is merely due to the fact that the PEP survey covers an area which is larger than those probed by surveys at smaller (mostly optical and near-IR) surveys. In fact, the above percentage rises to $\sim 95$\% in the inner portion of the GOODS-S, i.e. virtually all PEP sources are endowed with a redshift estimate.
 
Sources blindly detected at 100$\mu$m and 160$\mu$m above the 80\% completeness limit have then been used to calculate the angular two-point correlation function $w(\theta)$ of far-infrared galaxies over the whole redshift range spanned by the data. Under the assumption of a power-law behaviour $w(\theta)=A\theta^{1-\gamma}$, for a fixed value of $\gamma=1.8$, we find an amplitude $A\simeq 1.2\cdot 10^{-3} $, identical within the uncertainties at both frequencies. 
By using the redshift distributions derived from the MUSIC catalogue, we then obtain values for the comoving correlation length $r_0\sim 6.3$ Mpc  and $r_0\sim 6.7$ Mpc, respectively at 100$\mu$m and 160$\mu$m. Within the halo bias approach (Mo \& White 1996), such a clustering strength corresponds to dark matter haloes of masses $M\simgt10^{12.4} M_\odot$.

These values are in excellent agreement with those of Gilli et al. (2007), obtained for a sample of 
Luminous Infrared Galaxies selected at 24$\mu$m in the same GOODS South field.
 Also, the number densities of their sources and of our far-infrared galaxies coincide, indicating that these two classes of objects refer to the same galaxy 
population. \\
It is also worth noticing that our results are compatible with those derived at 250$\mu$m by Maddox et al. (2010) and Cooray et al. (2010) in the case of SPIRE-selected sources. In particular, Cooray et al. (2010) find values for the dark halo masses which are in excellent agreement with our estimates. 

With the aim of studying in greater detail the large-scale structure traced by actively star-forming galaxies in the redshift range where most of the cosmic star-formation activity is confined, we have then selected sources from the 100$\mu$m and 160$\mu$m catalogues in the $z=[1.7-2.6]$ interval. An analysis of the de-projected two-point correlation function in this case gives 
$r_0\sim 19$ Mpc and $r_0\sim 17.4$ Mpc at the two frequencies of interest.  By using once again the halo bias model, the above values correspond to halo masses $M\simgt10^{13.7} M_\odot$. \\
These are extremely high values and well match those found for similarly star-forming galaxies selected at other wavelenghts  in the same redshift range (e.g. Magliocchetti et al. 1999; Arnouts et al. 2002; Farrah et al. 2006b; Magliocchetti et al. 2007; 2008; Brodwin et al. 2008; Weiss et al. 2009; Foucaud et al. 2010; Hartley et al. 2010; Wake et al. 2011) and for quasars at earlier epochs (Shen et al. 2007).
Furthermore, by combining our clustering results with the information on stellar masses for galaxies in the GOODS-S, we find that the ratio 
between average values of the stellar mass $<M_*>$ at different epochs increases when going from $z\sim 1$ to $z\sim 2$ by about a factor 10 less 
than what happens in the case of the 
dark matter haloes hosting such galaxies ($M_*/M_{\rm min}(<z>\sim 1) \sim 4 \cdot 10^{-1}$ vs. $M_*/M_{\rm min}(<z>\sim 2) \sim 4 \cdot 10^{-2}$). 
When compared with recent results taken from the literature (e.g. Foucaud et al. 2010; Wake et al 2011; Moster et al. 2010 to mention a few), 
our findings converge at confirming the evolutionary picture of downsizing wherby massive galaxies at $z\sim 2$ were more actively forming stars than 
their $z\sim 1$ counterparts and at the same time also contained a lower fraction of their mass in (evolved) stars. 
This is also in agreement with the most recent results on the evolution of the cosmic star formation rate (cfr. e.g. Gruppioni et al. 2010) and 
indicates that the bulk of stellar formation within massive galaxies took place at $z\simgt 2$.
The intensity of this dramatic process then decreases as we approach the local universe, and already by $z\sim 1$ star forming activity 
is largely reduced as most of the stars within such galaxies have already been formed. 

A deeper investigation of the PEP maps indicate that the strong clustering signal detected for $z\sim 2$ PEP galaxies is due to the presence,  more visible at 100$\mu$m than in the other band, of a wide (at least 4 Mpc across in projection), $M\simgt 10^{14} M_\odot$,  filamentary structure which includes more than 50\% of the sources detected above the  $80$\% completeness level  at $z\sim 2$.  \\
Such an overdensity, already detected in Salimbeni et al. (2009) but with a lower statistical significance, extends all along the GOODS-S and seems to be made of a more diffuse structure at $z\sim 2.4$, 'bracketed' at the two ends of its length by two more concentrated clumps set at redshifts $z\sim 2.2$ and $z\sim 2$. The total number of 100$\mu$m-selected galaxies down to the PEP, 
$S_{100\mu\rm m}\simgt 1$ mJy flux limit and residing in the overdensity is 28. 
We stress that this sample is not expected to be complete, both for geometrical reasons (filament possibly extending beyond the edges of the surveyed area) and in terms of flux limits (completeness of the 100$\mu$m band just $\sim$10\% at $\sim$1 mJy).    


It is nevertheless interesting to compare some of the properties of the galaxies which reside within the overdensity with those of more isolated or 'field' sources, otherwise selected with the same criteria. 
For instance, a comparison between the distribution of bolometric luminosities for these two classes of objects shows a trend for galaxies belonging to the filamentary super-structure to be more luminous than field sources. If confirmed by analyses performed on larger datasets, this result also implies a higher star formation rate in proto-cluster objects as compared with isolated galaxies, finding which is likely due to the joint effect of large reservoirs of gas available at high redshifts in deep potential wells such as those associated to large overdensities and the enhanced rate of encounters between sources favoured by their proximity. 



As a last consideration, we note that extended proto-clusters of star-forming galaxies at $z\sim 2$ seem to be relatively common events, especially when found in association with radio sources (see e.g. Pentericci et al. 2000; Steidel et al. 2005; Andreani, Magliocchetti \& de Zotti, 2009; Hatch et al. 2010 and references therein). 
However, a systematic study of these objects is currently limited by the capabilities of the state-of-the-art instruments which either sample wide but shallow fields or concentrate on deeper/pencil-beam-like surveys.
Deeper/wider area surveys in the far-IR regime such as those planned for the SPICA mission (SPICA Study Team Collaboration, 2010) will be able to provide a 
more statistically significant census of these highly dense regions so to characterize the interplay between formation and evolution of the large-scale structure and of the galaxies that trace it right at the epoch when most of the cosmic action takes place.\\
\\
\noindent
{\bf Acknowledgements}
We thank the referee for his/her constructive suggestions which helped shaping and improving the paper.
PACS has been developed by a consortium of institutes led by MPE
(Germany) and including UVIE
(Austria); KU Leuven, CSL, IMEC (Belgium); CEA, LAM (France); MPIA
(Germany); INAF-
IFSI/OAA/OAP/OAT, LENS, SISSA (Italy); IAC (Spain). This development
has been supported by the
funding agencies BMVIT (Austria), ESA-PRODEX (Belgium), CEA/CNES
(France), DLR (Germany),
ASI/INAF (Italy), and CICYT/MCYT (Spain).

\end{document}